\newif\ifPDF
\ifx\pdfoutput\undefined\PDFfalse
\else\ifnum\pdfoutput > 0\PDFtrue
     \else\PDFfalse
     \fi 
\fi
\documentclass[useAMS,usenatbib]{mn2e}
\ifPDF
  \usepackage[T1]{fontenc}
  \usepackage{aeguill}
  \usepackage[pdftex]{graphicx,color}
  \usepackage{subfigure}
  \usepackage{afterpage}
\else
  \usepackage[T1]{fontenc}
  \usepackage[dvips]{graphicx,color}
  \usepackage{subfigure}
  \usepackage{afterpage}
\fi

\usepackage{graphics}
\usepackage{rotating}
\usepackage{multirow}

\def\zsun{Z$_{\odot}$}
\def\msun{M$_{\odot}$}

\def\mpc{\rm{Mpc$^{-1}$}}
\def\kms{\rm{km~s$^{-1}$}}
\def\cmq{\rm{cm$^{-3}$}}
\def\ergss{\rm{ergs\,s$^{-1}$}}

\def\te{$T_{\rm{e}}$}
\def\ne{$N_{\rm{e}}$}
\def\ebvg{E(\textit{B-V})$_{\rm gal}$}
\def\ebvt{E(\textit{B-V})$_{\rm tot}$}
\def\ebvi{E(\textit{B-V})$_{\rm int}$}

\def\lam{$\lambda$}
\def\two{\,{\sc ii\,}}
\def\three{\,{\sc iii\,}}
\def\four{\,{\sc iv\,}}

\newcommand{\ha}{H\ensuremath{\alpha}}
\newcommand{\hb}{H\ensuremath{\beta}}
\newcommand{\hg}{H\ensuremath{\gamma}}
\newcommand{\hd}{H\ensuremath{\delta}}
\newcommand{\mdot}{\ensuremath{\dot{M}}}

\def\hei{He\,{\sc i\,}}
\def\cii{C\,{\sc ii}\,}
\def\ciii{C\,{\sc iii}\,}
\def\civ{C\,{\sc iv}\,}

\def\nii{[N\,{\sc ii}]\,}
\def\niii{N\,{\sc iii}\,}
\def\niv{N\,{\sc iv}\,}
\def\nv{N\,{\sc v}\,}
\def\oi{[O\,{\sc i}]\,}
\def\oii{[O\,{\sc ii}]\,}
\def\oiii{[O\,{\sc iii}]\,}

\def\sii{[S\,{\sc ii}]\,}
\def\siii{[S\,{\sc iii}]\,}

\def\feiii{[Fe\,{\sc iii}]\,}

\def\ariii{[Ar\,{\sc iii}]\,}
\def\ariv{[Ar\,{\sc iv}]\,}
\def\heii{He\,{\sc ii}\,}
\def\siiv{Si\,{\sc iv}\,}
\def\sii{[S\,{\sc ii}]\,}
\def\cliii{[Cl\,{\sc iii}]\,}
\def\neiii{[Ne\,{\sc iii}]\,}

\newcommand{\h}{^{h}}
\newcommand{\m}{^{m}}
\newcommand{\hii}{H\textsc{ii}}
\newcommand{\eg}{{\rm e.g.}}
\newcommand{\ie}{{\rm i.e.}}

\newcommand{\cf}{{\rm cf}}
\newcommand{\mrtwo}{\multirow{2}{*}}

\newcommand{\mcone}{\multicolumn{1}{c}}
\newcommand{\mctwo}{\multicolumn{2}{c}}

\newcommand{\mcfive}{\multicolumn{5}{c}}
\newcommand{\mcsix}{\multicolumn{6}{c}}

\def\sb{{\sb}}

\title[The massive star population in Tol89] {The Massive Star
  Population in the Giant H\,{\Large\bf II} Region Tol\,89 in
  NGC\,5398\thanks{Based on observations collected at the European
    Southern Observatory, Chile, proposal ESO 73.B-0238(A) and with
    the NASA/ESA Hubble Space Telescope, obtained from the ESO/ST-ECF
    Science Archive Facility.}}

\author[F. Sidoli et al.]{Fabrizio~Sidoli$^{1}$\thanks{E-mail:
fs@star.ucl.ac.uk (FS)}, Linda~J.~Smith$^{1}$ 
and Paul~A.~Crowther$^{2}$\\
$^{1}$Department of Physics and Astronomy, University College London,
London WC1E 6BT, UK \\ 
$^{2}$Department of Physics and Astronomy,
University of Sheffield, Sheffield S3 7RH, UK}

\begin{document}

\date{}

\pagerange{\pageref{firstpage}--\pageref{lastpage}} \pubyear{2006}

\maketitle

\label{firstpage}

\begin{abstract}
  We present new high spectral resolution VLT/UVES spectroscopy and
  archival HST/STIS imaging and spectroscopy of the giant {\hii}
  region Tol\,89 in NGC\,5398. From optical and UV HST images, we find
  that the star-forming complex as a whole contains at least seven
  young compact massive clusters. We resolve the two brightest optical
  knots, A and B, into five individual young massive clusters along
  our slit, A1--4 and B1 respectively. From Starburst99
  (\citeauthor{leitherer}) UV spectral modelling, and nebular {\hb}
  equivalent widths in the optical, we derive ages that are consistent
  with the formation of two separate burst events, of $\sim4\pm1$ Myr
  and $<3$ Myr for knots A (A1--4) and B (B1), respectively.  An LMC
  metallicity is measured for both knots from a nebular line analysis,
  while nebular {\heii} 4686 is observed in knot B and perhaps in knot
  A.  We detect underlying broad wings on the strongest nebular
  emission lines indicating velocities up to 600 {\kms}.  From UV and
  optical spectroscopy, we estimate that there are $\sim$95 early WN
  stars and $\sim$35 early WC stars in Tol\,89-A, using empirical
  template spectra of LMC WR stars from \citeauthor{ch06}, with the WC
  population confined to cluster A2. Remarkably, we also detect a
  small number of approximately three mid WNs in the smallest (mass)
  cluster in Tol\,89-A, A4, whose spectral energy output in the UV is
  entirely dominated by the WN stars. From the strength of nebular
  {\hb}, we obtain N(O) $\sim$690 and 2800 for knots A and B,
  respectively, which implies N(WR)/N(O)$\sim$0.2 for knot A.  We also
  employ a complementary approach using Starburst99 models, in which
  the O star content is inferred from the stellar continuum, and the
  WR population is obtained from spectral synthesis of optical WR
  features using the grids from \citeauthor{snc02} We find reasonable
  agreement between the two methods for the O star content and the
  N(WR)/N(O) ratio but find that the WR subtype distribution is in
  error in the Starburst99 models, with far too few WN stars being
  predicted. We attribute this failure to the neglect of rotational
  mixing in evolutionary models.  Our various modelling approaches
  allow us to measure the cluster masses. We identify A1 as a super
  star cluster (SSC) candidate with a mass of $\sim1$--$2\times10^{5}$
  {\msun}. A total mass of $\sim6\times10^{5}$ {\msun} is inferred for
  the ionizing sources within Tol\,89-B.
  \end{abstract}


\section{Introduction}\label{introduction}

Giant {\hii} regions (GHRs) are characterised by their large sizes (up
to $\approx1$ kpc), supersonic gas motions \citep*{melnick99} and high
{\ha} luminosities
\citep[$10^{38}$--10$^{41}$\,ergs\,s$^{-1}$;][]{kennicutt}. A review
of their properties is given by \citet{shields90}. The presence of
GHRs in galaxies denotes sites of recent, intense episodes of massive
star formation. The nearest extragalactic GHR is 30 Doradus in the
Large Magellanic Cloud (LMC) which hosts a compact star cluster (R136)
of mass $\sim 2$--$6\times10^{4}$ M$_{\odot}$ \citep{hunter95} as the main
ionizing source.  Conversely, another nearby GHR, NGC 604 in M33, is
of similar size but contains multiple OB associations rather than a
central massive cluster. The mechanism(s) which ultimately determines
whether a cluster or a complex of OB associations is formed when an
intense star-forming event occurs is unclear. \citet*{elmegreen97}
suggest that clusters are formed in preference to loose associations
in high pressure interstellar environments. Clusters with masses as
high as $10^{5}$ to $\sim 10^{8}$ M$_{\odot}$ \citep{maraston04} --
often termed young massive clusters (YMCs), or super star clusters
(SSCs) -- are observed in the extreme environments of starburst
galaxies and galaxy mergers \citep[{\eg}][]{whitmore03}; the most
luminous and compact of which show similar properties to older
globular clusters (GCs) such as those observed in the Milky Way
\citep[{\eg}][see also \citeauthor{larsen04} 2004 for a
review]{holtzman92,whitmore97,schweizer98,bastian06}. This has led to
the suggestion that YMCs may represent the young counterparts of old
GCs \citep[{\eg}][]{ashmen92} and may offer insight into the formation
and evolution of GCs in the local universe.

Recently, \citet*{chen05} have studied the cluster content of three
GHRs (NGC\,5461, NGC\,5462 and NGC\,5471) in M\,101. They find that
they contain clusters similar to R136 in mass rather than the more
massive SSCs, although they note that the three R136-like clusters in
NGC 5461 may merge to form an SSC.  Overall, they find evidence for a
link between the molecular cloud distribution and the cluster
luminosity function in the sense that a diffuse distribution produces
more lower mass clusters, in support of the hypothesis that massive
clusters are formed in high pressure environments.

\begin{figure}
\begin{minipage}[t]{8.4cm}
\includegraphics[scale=0.43,angle=-90,clip=true]
{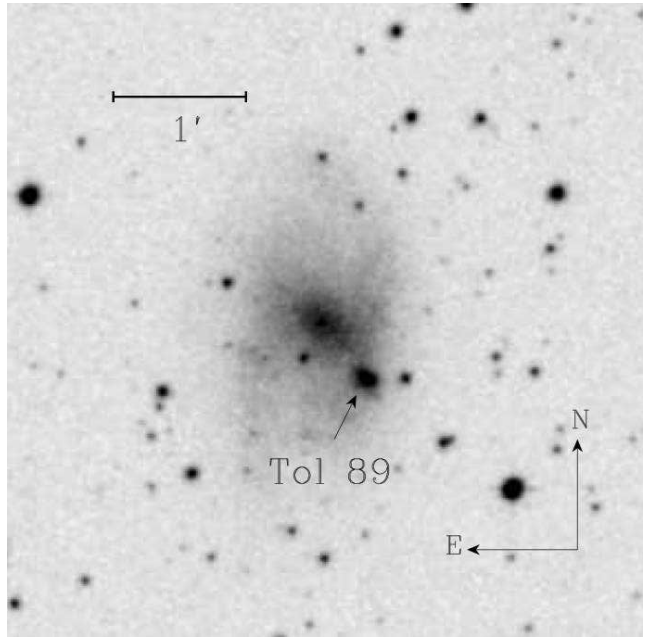}
\caption{Digital Sky Survey R-band image of NGC\,5398 showing 
  the location of the GHR Tol\,89. At the adopted distance of 14.7
  Mpc, $1'$ corresponds to $\approx$\, 4\,kpc. The field of view is $5\times5$
  arc minutes.}\label{ngc5398}
\end{minipage}
\end{figure}

The giant {\hii} region Tol\,89 \citep{smith76} is located at the
south-western end of the bar in the late-type barred spiral (Sdm)
galaxy NGC\,5398 \citep{durret}. A Digital Sky Survey image of this
galaxy is shown in Fig.~\ref{ngc5398}; Tol\,89 is conspicuous in being
the only large massive star-forming complex present in the entire
galaxy. Tol\,89 has an extent of $\approx24''\times18''$ or
$1.7\times1.2$ kpc and an absolute blue magnitude of $-14.8$,
\citep[assuming a distance of 14.7\,Mpc based on $H_{0}=75$
{\kms}{\mpc};][]{schaerer99} which makes it one of the most impressive
GHRs known.

The presence of Wolf-Rayet (WR) stars in Tol\,89 was first reported by
\citet{durret} who detected a broad emission bump at \lam4650.
\citet*{schaerer} identified broad features of N{\three}4640,
He{\two}4686, and very strong C{\four}5808, suggesting the presence of
both late-type WN (WNL) and early-type WC (WCE) stars in Tol\,89. From
a detailed spatial analysis, they find an offset between the continuum
and nebular lines, suggesting a complex star-forming region.
\citet*{johnson03} obtained radio observations of Tol\,89 and
discovered an unresolved thermal radio source with a Lyman continuum
flux of $\sim4500\times 10^{49}$\,s$^{-1}$, equivalent to an SSC with
a mass of $10^{6}$\,M$_{\odot}$ if the source is a single cluster. In
terms of radio luminosity, they find that Tol\,89 is among the most
luminous radio {\hii} regions so far observed and is comparable to the
GHR NGC\,5471 in M\,101.  Finally, Tol\,89 was one of the objects in
the survey of \citet{chandar04} to measure the WR content of actively
star-forming regions through ultraviolet (UV) spectroscopy obtained
with the {\it Hubble Space Telescope} (HST) Space Telescope Imaging
Spectrograph (STIS).

In this paper, we examine the massive stellar content of Tol\,89
through an analysis of archival HST images and UV spectroscopy, and
high resolution optical spectroscopy obtained with the {\it Very Large
  Telescope} (VLT). We show that Tol\,89 is a young, very massive
star-forming complex with at least seven young compact massive
clusters.  The fact that Tol\,89 is located at the end of the bar in
NGC\,5398 indicates it may have been formed through gas inflow in a
high pressure environment, although \cite{johnson03} suggest that the
weak bars found in late-type galaxies are not strong enough to
generate the required gas inflow.

The paper is structured as follows. The observations and reductions
are presented in Section~\ref{observations}. In Section~\ref{spectra}
we describe the spectra and in Sections~\ref{tol89_properties} and
\ref{nebular_properties} we derive the properties of the knots and
their ionizing clusters. In Section~\ref{mpop} we estimate the massive
star content using empirical and synthesis techniques.  Finally, in
Sections~\ref{discussion} and \ref{summary} we discuss and summarise
our results.

\section{Observations}\label{observations}

We have obtained high spectral resolution \textit{UV-Visual Echelle
  Spectrograph} (UVES)+VLT echelle spectroscopy of the brightest
optical knots in the giant {\hii} region Tol\,89. We supplement this
dataset with archival HST imaging and spectroscopy obtained using the
STIS CCD and FUV-MAMA detectors in the optical and UV respectively
(Proposal ID 7513; C.~Leitherer, P.I.). The UV spectroscopic data are also
presented in \citet{chandar04}. A summary of the observations is given
in Table~\ref{obs}. All the HST archive data were processed using
the standard {\sc calstis} data reduction pipeline.

\subsection{Imaging}\label{imaging}

\begin{figure*}
\begin{minipage}[t]{8.6cm}
\includegraphics[scale=0.44,angle=-180,clip=true]
{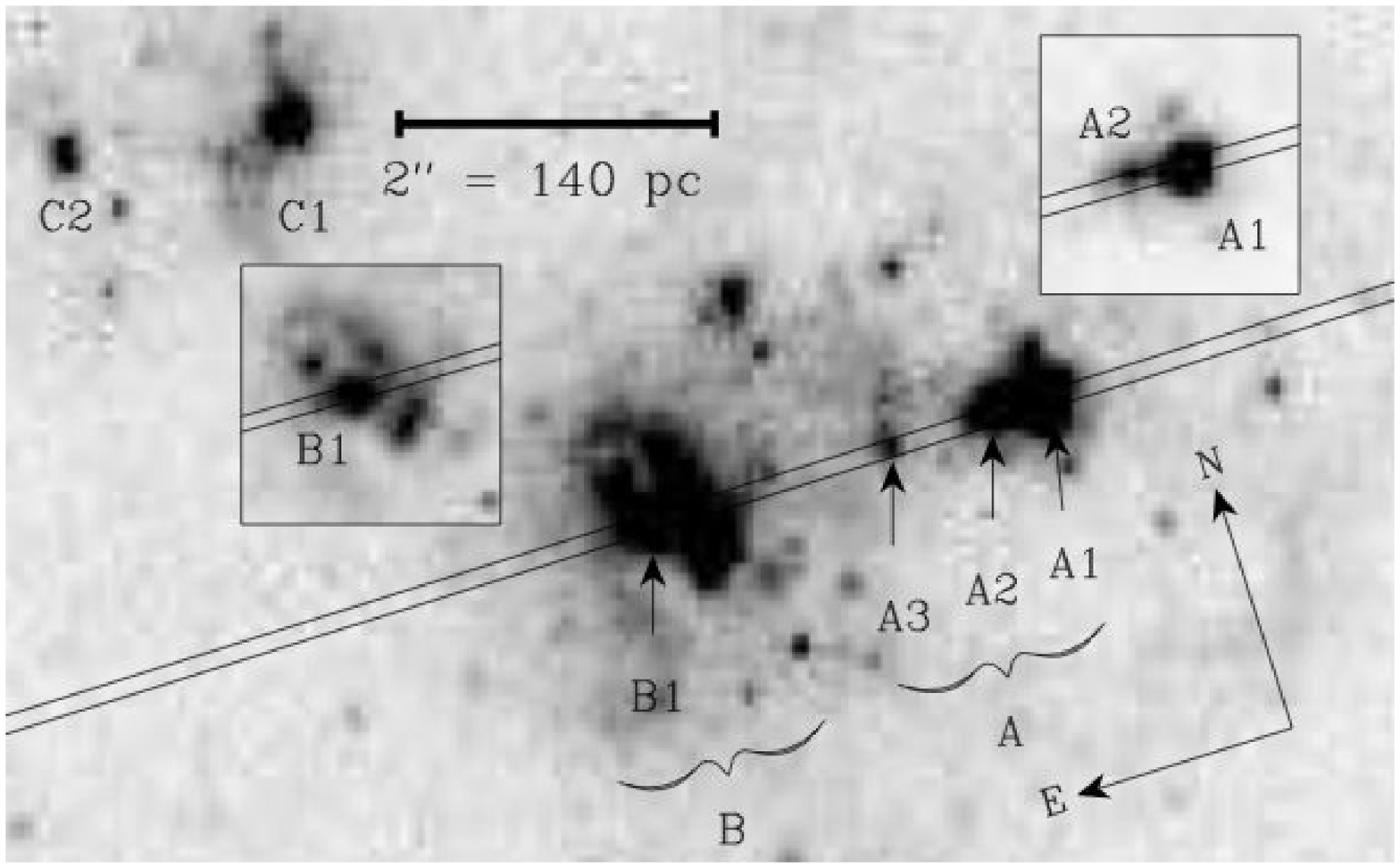}
\caption{\label{stis_vis} STIS MIRVIS ($\lambda_{\rm{cen}}$=7230
  {\AA}) image showing the position and orientation
  (PA\,=\,-90{\degr}) of the optical STIS slit, of size
  $52''\times0\farcs1$ arc seconds. Clusters A1-3, B1, C1 and C2 are
  indicated (see Section~\ref{morphology}). At the adopted distance of
  14.7 Mpc, $1''$ corresponds to $\approx70$ pc. The stretch of the
  image has been set so as to highlight the position of cluster A3.
  The insets show the regions around A1 and A2 (top right) and B1
  (centre left) at different stretches to highlight better these
  clusters.}
\end{minipage}
\hspace{0.3cm}
\begin{minipage}[t]{8.6cm}
\includegraphics[scale=0.44,angle=-180,clip=true]
{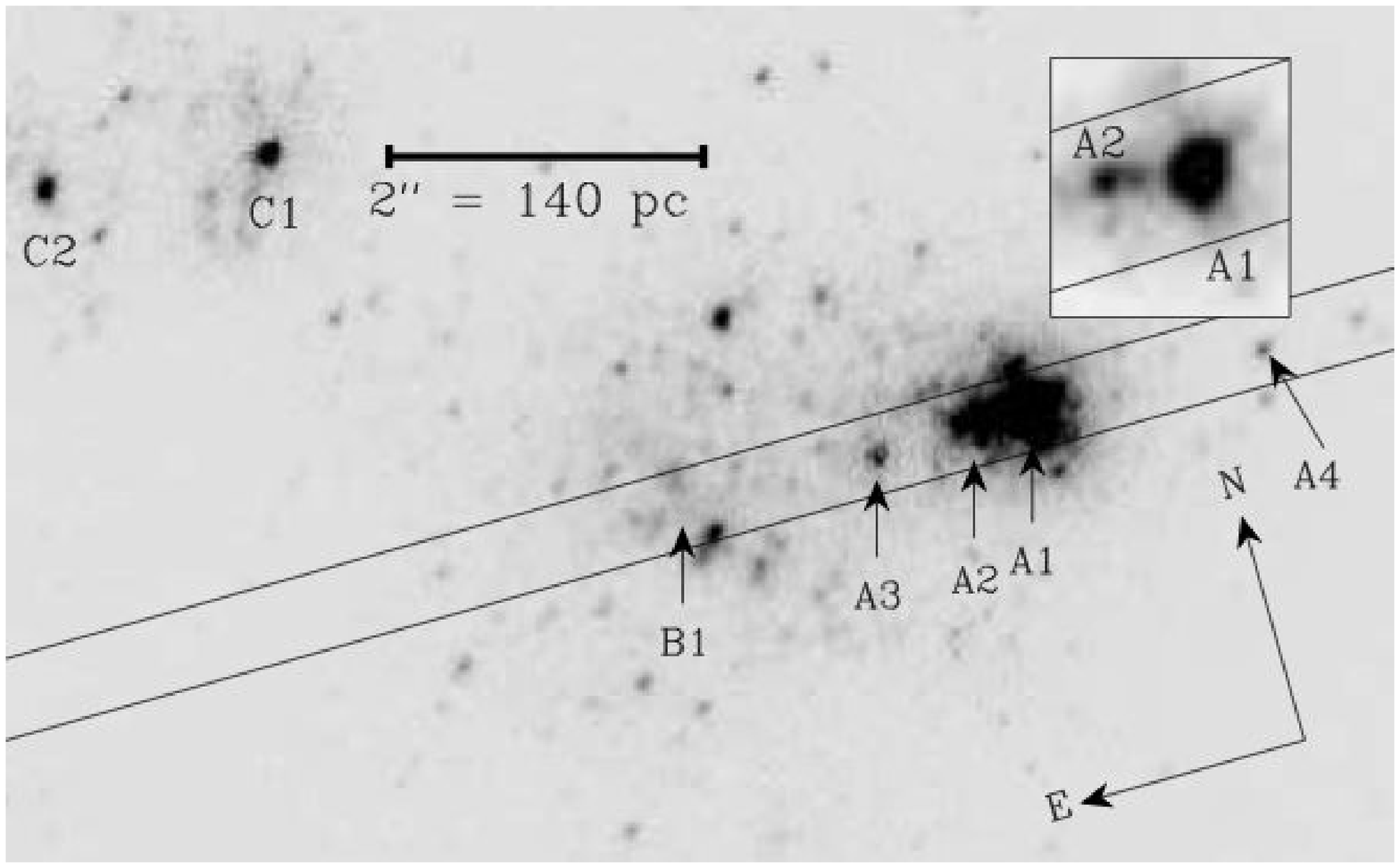}
\caption{\label{stis_uv}STIS MIRFUV ($\lambda_{\rm{cen}}$=1480 {\AA})
  image showing the position and orientation (PA\,=\,-90{\degr}) of
  the UV STIS slit, of size $52''\times0\farcs5$. Clusters A1-4, B1,
  C1 and C2 are indicated (see Section~\ref{morphology}). At the
  adopted distance of 14.7 Mpc, $1''$ corresponds to $\approx70$ pc.
  The stretch of the image has been set so as to highlight the
  position of clusters A3 and A4. The inset shows the region around
  clusters A1 and A2. Note, cluster B1 is almost completely obscured
  in the UV.}
\end{minipage}
\end{figure*}

\begin{figure}
\begin{minipage}[t]{8.6cm}
\includegraphics[scale=0.44,angle=-180,clip=true]
{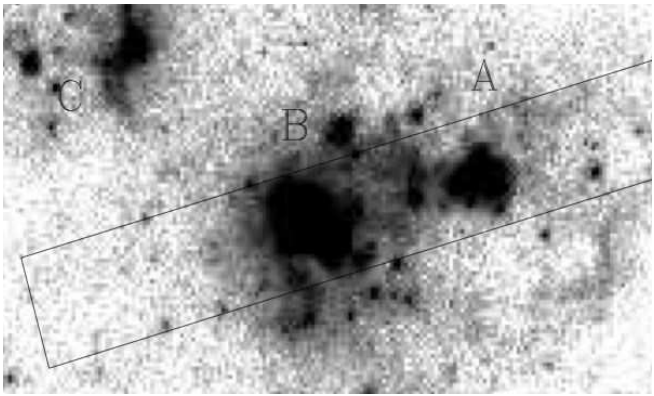} 
\caption{\label{uves_vis} STIS MIRVIS ($\lambda_{\rm{cen}}$=7230 {\AA})
  image showing the position and orientation (PA\,=\,90{\degr}) of the
  UVES slit, of size $1\farcs4\times10''$. Knots A, B and C are
  indicated (see Section~\ref{morphology}). The orientation and scale
  are as in Figs.~\ref{stis_vis} and \ref{stis_uv}.}
\end{minipage}
\end{figure}

\subsubsection{Reduction}\label{image_reduction}

Optical and UV images were obtained on 1999 February 19 using the
MIRVIS/STIS CCD and MIRFUV/FUV-MAMA longpass filters/detectors. The
optical dataset consists of four image sets. Before combining, we
first used intensity histograms of the images to determine a threshold
value for the cold pixels, below which all pixels were flipped to high
values. These were subsequently removed, along with cosmic rays and
hot pixels, using the \textsc{iraf/stsdas} cosmic ray rejection
(\textsc{occrej}) task upon combining the images. The resulting images
were then geometrically corrected using the \textsc{x2d} task.
Remaining hot pixels were removed using the Laplacian cosmic ray
identification algorithm, L.A.Cosmic \citep{van_dokkum}. In
Figs.~\ref{stis_vis} and \ref{stis_uv} we show the optical and UV STIS
images of Tol\,89, with the slit positions of the STIS spectroscopy
superimposed (see Section~\ref{stis_spectroscopy}). Similarly in
Fig.~\ref{uves_vis} we show the optical STIS image with the slit
position of the UVES spectroscopy overlaid
(Section~\ref{uves_spectroscopy}).

\subsubsection{Optical and UV morphology}\label{morphology}

Tol\,89 is a young massive star forming complex as shown by the
optical and UV STIS images in Figs.~\ref{stis_vis} and \ref{stis_uv}.
It comprises three distinct knots of star formation that we label
A, B and C in order of decreasing (optical) brightness. With the high
spatial resolution of HST, we are able to resolve each individual knot
into multiple cluster components.

Along the optical STIS slit, the boundary between A and B is defined
by the mid-point between the brightest optical component in each knot.
We then identify the clusters in order of decreasing brightness along
the slit, with 1 being the brightest (see Figs.~\ref{stis_vis} and
\ref{stis_uv}).  The UV counterparts to each cluster were then
identified in the FUV-MAMA image. The separation between the clusters,
as measured from the optical STIS image, are as follows:
A4$\rightarrow$A1: 1\farcs53 (107 pc); A1$\rightarrow$A2: 0\farcs32
(22 pc); A2$\rightarrow$A3: 0\farcs69 (49 pc); and A3$\rightarrow$B1:
1\farcs51 (106 pc). The projected separations in parsecs are given in
parentheses, where 1$'' \approx70$ pc \citep[assuming a distance of
14.7\,Mpc based on $H_{0}=75$ {\kms}{\mpc};][]{schaerer99}.

Knot A is made up of multiple clusters that are bright in both the
optical and the UV, and dominated by a young compact cluster which we
denote as A1. Whilst knot B is also optically bright and complex it is
much fainter than A in the UV. It is also the location of the
impressive thermal radio source discovered by \citet{johnson03}. From
the STIS UV image (Fig.~\ref{stis_uv}) we can see that knot C
contains two young compact massive clusters which we denote as C1 and
C2; where C1 is the location of the north-eastern spur of thermal
radio emission identified by \citealt{johnson03} (see their fig.~4).

\citet{durret} identify two regions in Tol\,89 that they denote as X
and C, where C is their centre of Tol\,89. We identify our knot C with
their region X and their centre of Tol\,89 with our knots A and B.
Within the centre of Tol\,89, \citet{schaerer99} identify an area with
maximum nebular intensity (\,=\,our B), offset by $\sim2$ arc seconds to
the east of the region with maximum continuum intensity and the
location of the WR emission (\,=\,our A; see also their
fig.~5\footnote{Note that the orientation in fig.\,5 of
  \citeauthor{schaerer} 1999 is incorrect. North points towards the
  top right corner of their middle panel rather than the top left.
  Thus, the peak in nebular intensity (our knot B) is offset by
  $\sim2$ arc seconds to the east of the continuum intensity peak (our
  knot A) rather than south, as is suggested in their figure.}).

\begin{table*}
\caption{Summary of the observations. The HST+STIS imaging and 
spectroscopy (Proposal ID 7513, C.~Leitherer, P.I.) have been obtained 
from the HST archive.}\label{obs}
\setlength{\tabcolsep}{3.95mm}
\begin{tabular}{|l|c|c|c|c|c|c|c|c|}\hline\hline

\mrtwo{Detector} & \mrtwo{Filter}  & $\lambda_{\rm{cen}}$ 
 & FWHM & PA & Aperture Size & Pixel Scale & Total Exp. (s)/ \\ 

& & ({\AA}) 
& ({\AA})  & ({\degr}) & ($''$) & ($''$pix$^{-1}$) & Imsets \\ 
\hline

\\
\multicolumn{8}{c}{HST+STIS Imaging, 1999 February 19} \\
\\
FUV-MAMA &  MIRFUV   & 1480 & 280  & 106.7 & 25$\times$25 & 0.0246 & 900/1 \\
CCD      &  MIRVIS~~ & 7230 & 2720 & 106.7 & 28$\times$50 & 0.0507 & 240/4 \\
\\

\end{tabular}

\setlength{\tabcolsep}{1.2mm}
\begin{tabular}{|l|l|c|c|c|c|c|c|c|c|}
\hline\hline\\
\mrtwo{Detector} & Grating/ & $\lambda_{\rm{cen}}$ 
& $\lambda_{\rm{range}}$ & Resolution & Dispersion 
& PA & Slit Size & Pixel Scale & Total Exp. (s)/ \\ 

& Setting & ({\AA}) & ({\AA}) & ({\AA}) 
& ({\AA}pix$^{-1}$) & ({\degr}) & ($''$) 
& ($''$pix$^{-1}$) & Imsets \\ \hline

\\
\multicolumn{10}{c}{HST+STIS Spectroscopy, 2000 March 20-21} \\
\\

FUV-MAMA    & G140L & 1500  & 1150-1700 & 3.1 & 0.58  
& 89.0 & 52$\times$0.5 & 0.0246 & 11620/4 \\

CCD         & G430L & 3200  & 2900-5700 & 4.9 & 2.73  
& 89.1 & 52$\times$0.1 & 0.0507 & 810/2   \\

CCD         & G750M & 6581  & 6295-6867 & 1.0 & 0.56  
& 89.1 & 52$\times$0.1 & 0.0507 & 825/2   \\

\\
\multicolumn{10}{c}{UVES+VLT Spectroscopy, 2004 May 6} \\
\\
EEV CCD & Dichroic \#1 & 3460 & 3030-3880  & 0.19            
& 0.035 & 90.0 & 1.4$\times$10 & 0.246 & 2850/2 \\

EEV CCD & Dichroic \#2 & 4370 & 3730-4990  & 0.22            
& 0.044 & 90.0 & 1.4$\times$10 & 0.246 & 2850/2 \\

EEV/MIT-LL CCD & Dichroic \#1 & 5640 & 4580-6680  & 0.36/0.29$^{a}$ 
& 0.049/0.040$^{a}$ & 90.0 & 1.4$\times$11 & 0.182 & 2850/2 \\

EEV/MIT-LL CCD & Dichroic \#2 & 8600 & 6660-10600\,\,\, &0.63/0.45$^{a}$ 
& 0.075/0.060$^{a}$ & 90.0 & 1.4$\times$12 & 0.170 & 2850/2 \\ \hline

\end{tabular}

\medskip
$^{a}$ Resolutions and dispersions are for CCD1/CCD2\hspace{9.9cm}

\end{table*}

\subsection{Spectroscopy}\label{spectroscopy}

\subsubsection{STIS spectroscopy}\label{stis_spectroscopy}

The two-dimensional STIS spectroscopic data were obtained over two
visits on 2000 March 20-21 using the G430L and G750M gratings and the
STIS CCD detector for the optical regime, and the G140L grating and
the STIS FUV-MAMA detector for the UV. The slits were centred on the
brightest cluster (=A1) at co-ordinates of $\alpha=14\h01\m19\fs92$;
$\delta=-33\h04\m10\fs7$ (J2000). The respective slit sizes are
$52''\times0\farcs1$ and $52''\times0\farcs5$ and are shown in
Figs.~\ref{stis_vis} and \ref{stis_uv}. Further details are given in
Table~\ref{obs}. The image sets (two for each optical grating and four
for the UV) were combined using the method described in
Section~\ref{image_reduction} and rectified, wavelength and absolutely
flux calibrated using the \textsc{x2d} task.
From measurements of the FWHM of Gaussian fits to the unresolved {\hb}
and {\cii}1335 lines in the G430L and G140L spectra of B1, we obtain
spectral resolutions of 4.9{\AA} and 3.1{\AA} respectively, or 1.8
pixels, giving a resolution of 1.0{\AA} for the G750M grating.

From the UV spectral image, we identify five regions for extraction
which we denote as A1, A2, A3, A4 and B1 in Fig.~\ref{stis_vis} (see
Section~\ref{morphology}). The extraction widths correspond to 0.23,
0.12, 0.19, 0.19 and 0.30 arc seconds, respectively. Note that in
Fig.~\ref{stis_uv} the bright UV source in the vicinity of B1 is
actually the UV counterpart to the optical point source that lies
below, and just outside of the narrower optical slit (see leftmost
inset of Fig.~\ref{stis_vis}). For this reason we make careful
considerations regarding the extraction width of B1 to ensure that
there is as little contamination from this source as possible.

For the optical we extract three regions A1+2, A3 and B1, with the
extraction widths being defined by the full width at $\sim$10 per cent
of the peak of the intensity profile at {\ha}. These extraction
widths, corresponding to 0.55, 0.40 and 1.13 arc seconds respectively,
were then applied to the G430L grating spectra. Note, regions A1 and
A2 are barely resolved in the optical STIS 2D spectral image and hence
are extracted as a single source object while cluster A4 is not
detected above the background noise. Background subtraction was
performed by selecting regions of sky free of nebular emission. No
correction for slit losses have been made.

\subsubsection{UVES spectroscopy}\label{uves_spectroscopy}

Echelle spectra of Tol 89 were obtained in service mode on 2004 May 6
with UVES \citep{DOdorico} at the VLT Kueyen Telescope (UT2) in Chile
(Proposal ID 73.B-0238A, L.~J.~Smith, P.I.). UVES is a two arm
cross-dispersed echelle spectrograph with the red arm containing a
mosaic of an EEV and a MIT-LL CCD. The blue arm has a single EEV CCD;
all three CCDs have a pixel size of 15$\mu$m.  Simultaneous
observations in the blue and the red were made using the standard
setups with dichroic \#1 (346+564~nm) and dichroic \#2 (437+860~nm),
covering an almost continuous wavelength region from 3100--10\,360{\AA};
the regions between 5610--5670{\AA} and 8540--8650{\AA} were not
observed as a result of the gap between the two CCDs in the red arm.
The slit was oriented at a position angle of 90 deg to pass through
the two brightest knots in Tol 89, hereafter A and B when discussing
the UVES data (see Fig.~\ref{uves_vis}). The slit width was set at
1\farcs4, giving a resolving power of $\sim30\,000$ in the blue and
$\sim28\,000$ in the red. The pixel scales for $1\times1$ binning are
shown in Table~\ref{obs}.

At the time of the observations the conditions were clear and the
seeing was typically better than 1\farcs2.  A total integration time of
47.5 min. was divided into two equal length exposures to prevent
saturation of the brightest nebular lines ({\ha}, {\hb} and the
{\oiii}$\lambda\lambda4959,5007$). Observations of the
spectrophotometric standard Feige 67 were also made using the same set
up for the purpose of performing flux calibrations. Spectra of Th-Ar
comparison arcs were obtained to perform wavelength calibrations.

The data were reduced according to the standard CCD and echelle
reduction procedures within \textsc{iraf} using the \textsc{ccdred}
and \textsc{echelle} packages. This included the subtraction of a bias
level determined from the overscan region, bias subtraction, division
by a normalised flat field and bad pixel correction. Images were
cleaned of cosmic rays using L.A.Cosmic \citep{van_dokkum}. The
echelle orders were extracted using widths set to 2.4 and 3.6
arc seconds for knots A and B respectively and wavelength calibrated,
atmospheric extinction corrected and flux calibrated. Resolutions -- as
measured from the FWHM of Gaussian profile fits to the Th-Ar arc
lines -- and dispersions are given in Table~\ref{obs}.

\section{Description of the Spectra}\label{spectra}

\subsection{STIS UV}\label{stis-uv_spectra}

\begin{figure}
\includegraphics[scale=0.6,angle=0,clip=true]
{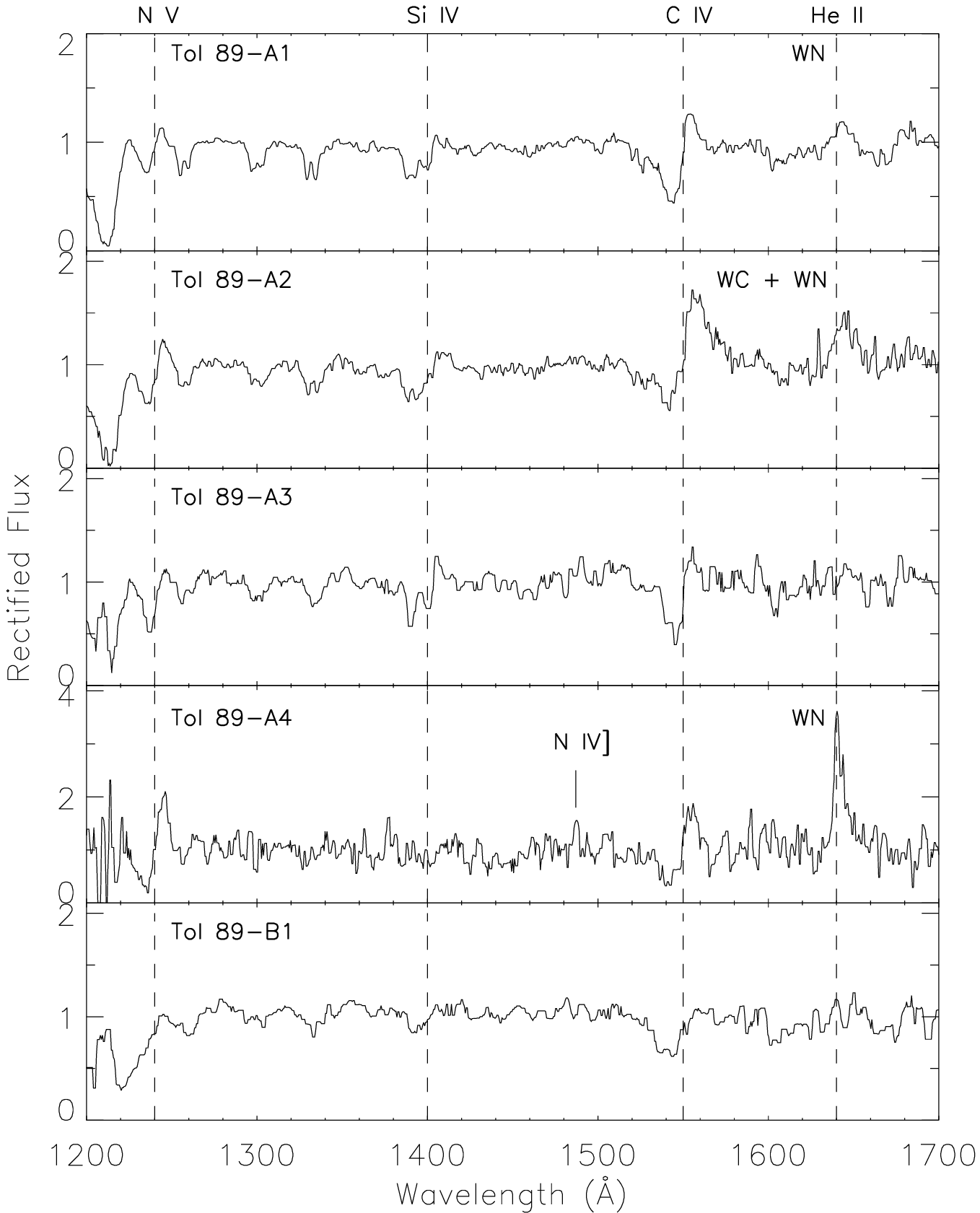}
\caption{\label{uv_spectra} Velocity corrected STIS FUV-MAMA G140L
  spectra showing the extractions of cluster A1, A2, A3, A4 and B1
  (\textit{top, 2nd, 3rd, 4th} and \textit{bottom} panels
  respectively). P Cygni profiles due to the winds of massive O and WN
  stars are identified: {\siiv}1400 (O supergiants), {\nv}1240 and
  {\civ}1550 (O stars) and {\heii}1640 (WN). In the case of A2, the
  strength of the {\civ}1550 feature, which exceeds that of
  {\heii}1640, suggests the presence of WC stars.}
\end{figure}

The velocity corrected ($V_{\rm{hel}}\approx$ 1230--1240 {\kms}; see
Section~\ref{nebular_properties}) UV spectra of clusters A1--4 and B1
are shown in Fig.~\ref{uv_spectra}; the characteristics of stars with
strong stellar winds due to the presence of P Cygni profiles of
{\siiv}1400, {\nv}1240 and {\civ}1550 can be seen. In the case of A3,
the narrow and resolved {\siiv}1400 profile is suggestive of late-type
O supergiants.

Since O stars do not show strong {\heii} 1640 emission, its presence
in A1, A2 and A4 can be attributed to the winds from WN stars. In A2
the emission strength of the {\civ}1550 feature, which is greater than
{\heii}1640, suggests that WC stars are also present, while the non
detection of {\heii} 1640 emission in A3 and B1 would suggest an
absence of WRs. In cluster A4 we appear to detect N\,{\sc iv}]\, 1486
emission from mid-WN stars and note that its UV spectral appearance
bares a close resemblance to that of a mid-WN star \citep[see fig.~2
of][]{cd98}.

The UV STIS data are also presented in fig. 1 of \citet{chandar04},
who used a larger aperture width of 81\,pixels to obtain their 1D
spectrum of the region they denote as Tol\,89-1. Their extraction
comprises our regions A1 and A2, plus diffuse inter-cluster emission
and also shows the P Cygni profiles of {\nv}1240, {\siiv}1400, and
{\civ}1550, as well as {\heii}1640 emission from WN stars. Our
smaller, individual extractions of A1 and A2 ($\sim$ 9 and 5 pixels
respectively) have allowed for different WR populations (WN and WC) to
be identified in the individual clusters. From the spectrum of
Tol\,89-1 shown in fig. 1 of \citeauthor{chandar04} the presence of WC
stars is ambiguous and thus only a WN population can be inferred from
their extraction.

\subsection{STIS optical}\label{stis-optical_spectra}

\begin{figure}
\includegraphics[scale=0.61,angle=0,clip=true]
{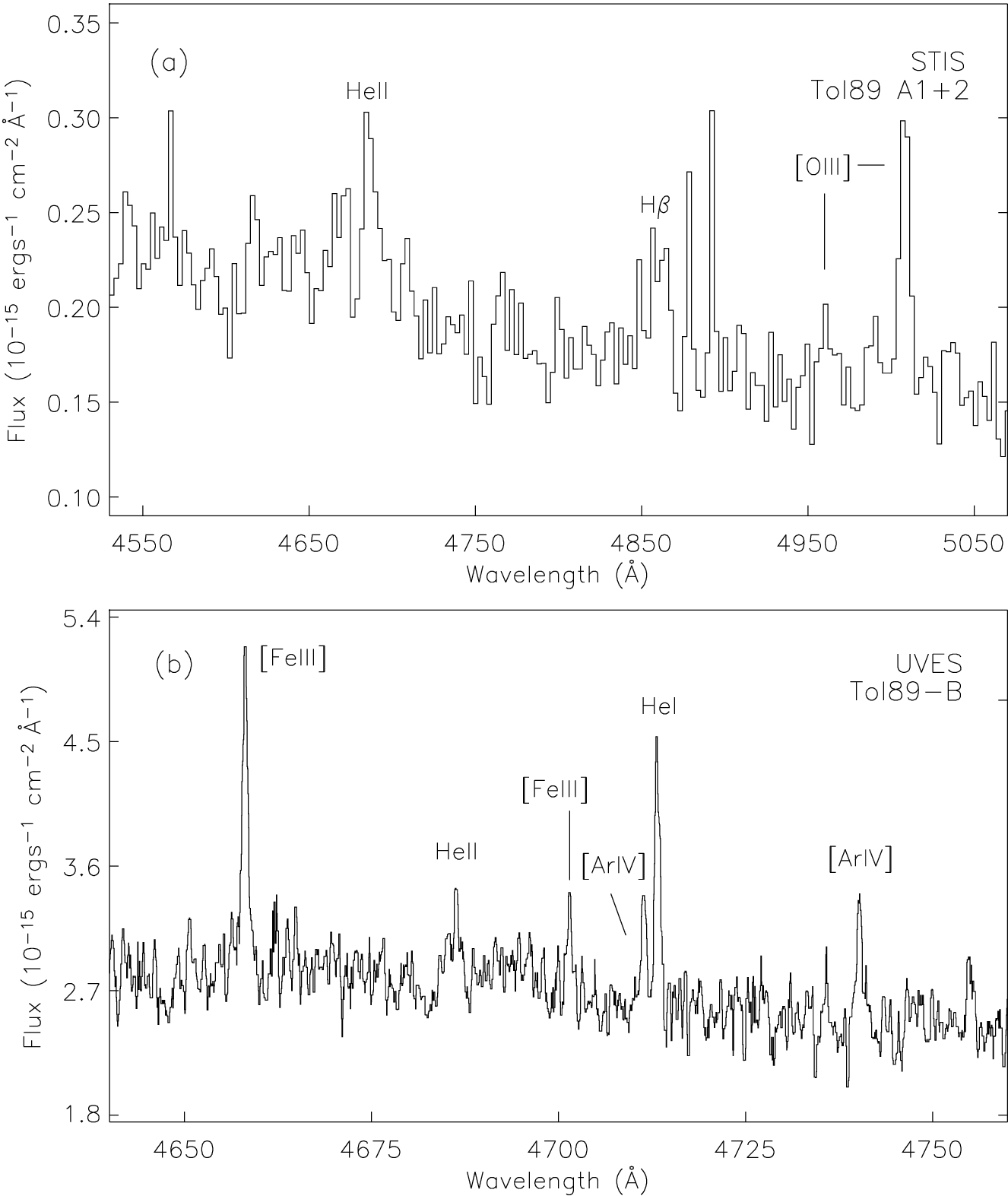} 
\caption{\label{nebheii} Velocity corrected STIS G430L spectrum of
  A1+2 (\textit{top}) and UVES+VLT spectrum of B (\textit{bottom})
  showing possible \textit{nebular} {\heii}4686 emission. Nebular
  lines of {\hb\, and \oiii} in A1+A2, and {\hei, \feiii\, and \ariv} in
  region B are also identified.}
\end{figure}

The optical spectrum of A1+2 shows a very blue continuum with very
weak emission lines of {\ha}, {\hb}, {\oiii}4959, 5007 and
{\nii}6548/84. The portion of the G430L spectrum in the region of
{\heii}4686 is shown in Fig.~\ref{nebheii}a; we identify a possible
nebular {\heii}4686 component. The FWHM is comparable to that of the
unresolved {\oiii}5007 emission line ({\cf} 7.0$\pm$1.8 and
6.7$\pm$1.0 {\AA} respectively). We briefly discuss the origin of this
emission in Section~\ref{discussion}.

Clusters A3 and B1 also show emission lines of {\ha}, {\hb} and
{\oiii}4959, 5007; although they are weak in A3. In addition to
these lines, B1 shows weak emission lines of {\oii}3727, {\hg} and
{\hd}. Both A3 and B1 are significantly reddened compared to A1+2; no
WR emission features are detected, in agreement with the UV STIS
spectra (see Fig.~\ref{uv_spectra}).

\subsection{UVES}\label{uves_spectra}

\begin{figure*}
\includegraphics[scale=0.9,angle=0,clip=true]
{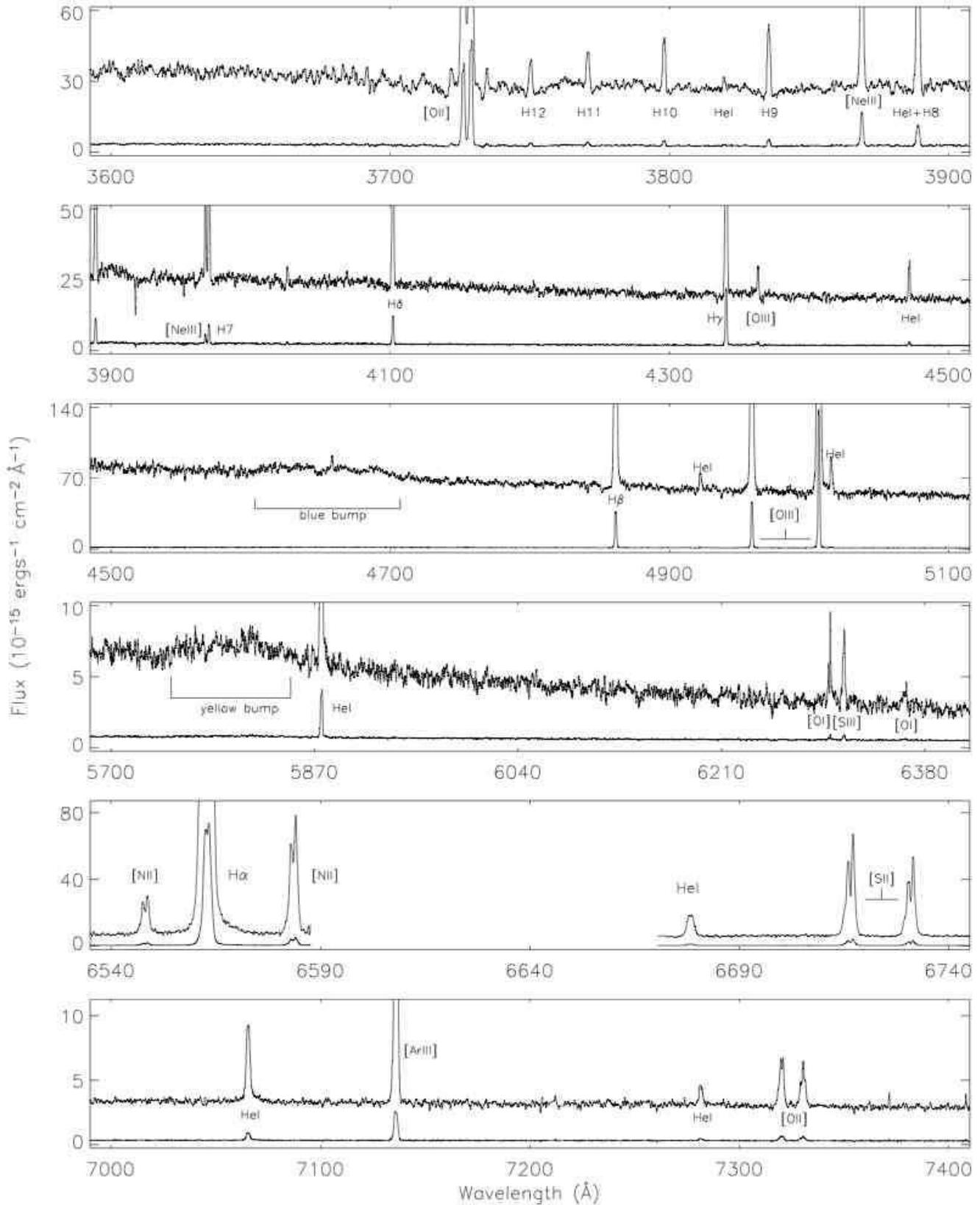} 
\caption{\label{uves_a} Velocity corrected UVES+VLT spectra of
  Tol\,89-A covering the spectral range of 3600--7400 {\AA}. In each
  panel the spectra are scaled to arbitrary flux values to show the
  detail in the spectra. The unscaled spectra are plotted to show the
  relative line intensities of the strongest emission lines. The
  velocity splitting in the strongest emission lines can be seen. The
  blue and yellow WR bumps are also indicated.}
\end{figure*}

\begin{figure*}
\includegraphics[scale=0.9,angle=0,clip=true]
{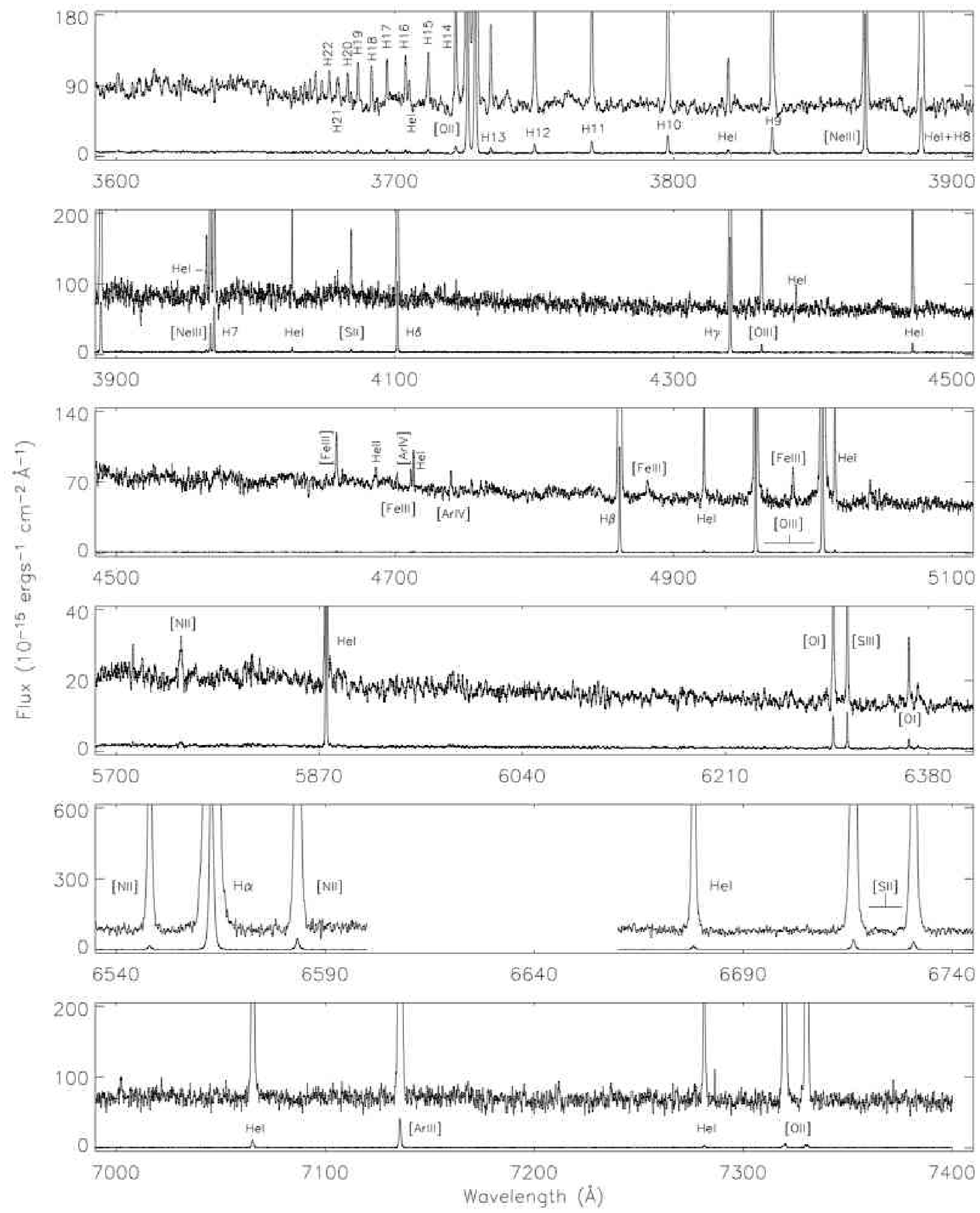}
\caption{\label{uves_b} Velocity corrected UVES+VLT spectra of
  Tol\,89-B covering the spectral range of 3600--7400 {\AA}. In each
  panel the spectra are scaled to arbitrary flux values to show the
  detail in the spectra. The unscaled spectra are plotted to show the
  relative line intensities of the strongest emission lines.}
\end{figure*}

The high S/N, high spectral resolution UVES data for regions A and B
are shown in Figs.~\ref{uves_a} and \ref{uves_b} respectively over the
wavelength range of 3600--7400 {\AA}. The spectra are rich in emission
line features typical of {\hii} regions. The strongest emission lines,
those of {\ha}, {\hb}, {\oii}3726/29, 7319/30, {\oiii}4959, 5007,
{\nii}6548/84, and {\sii}6716/31, exhibit a broad component resulting
in extended wings. In the case of region A, the narrow components of
these emission lines are split into two velocity components (denoted
as V1 and V2). Both spectra show little or no stellar absorption in
the Balmer lines.  Metal lines due to an older population of stars are
also absent and imply that regions A and B consist of a young
population of stars with very similar ages. An analysis of the nebular
lines is presented in Section~\ref{nebular_properties}.

\begin{figure}
\includegraphics[scale=0.63,angle=0,clip=true]
{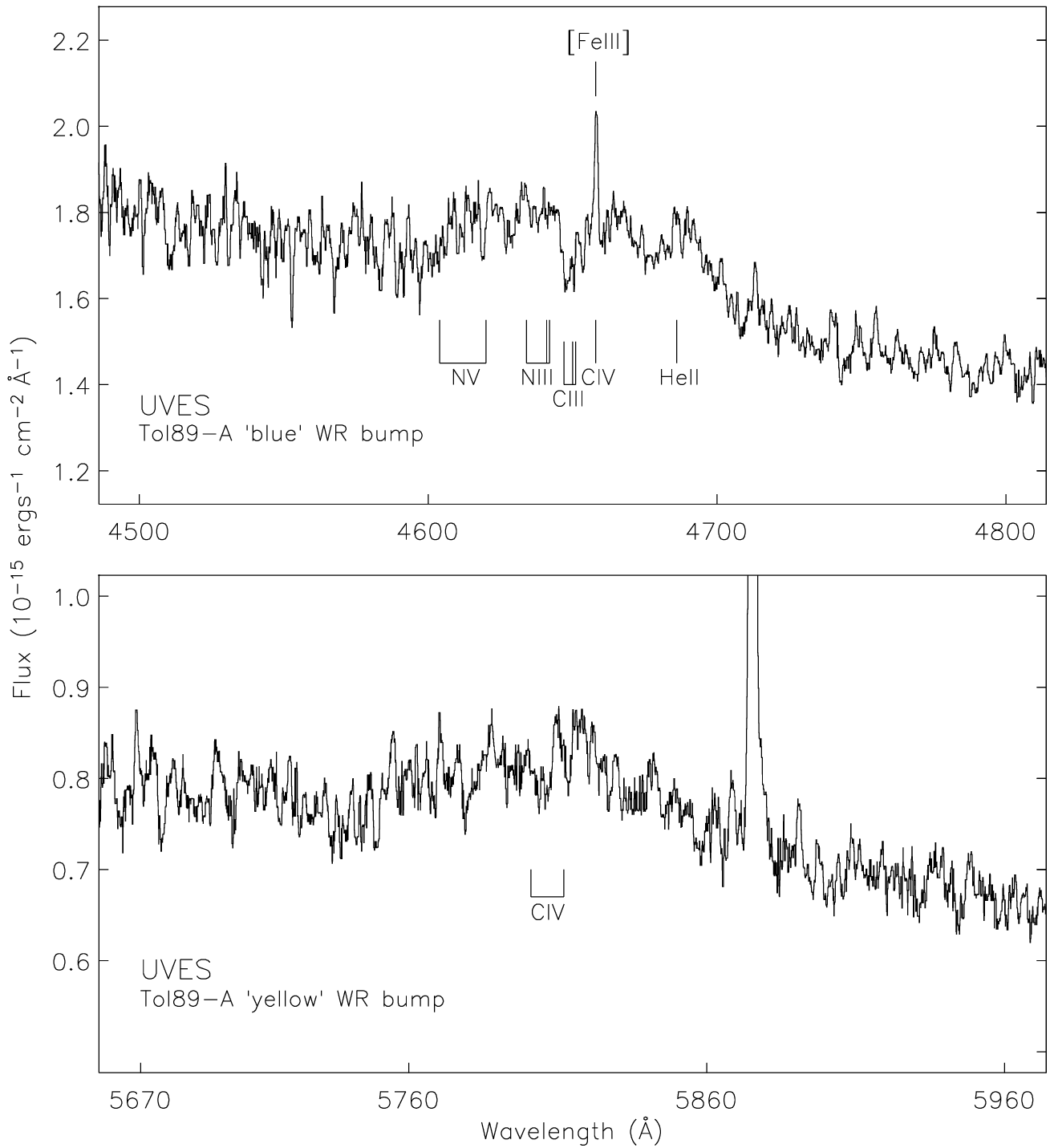}
\caption{\label{awrbumps} Velocity corrected UVES+VLT spectra showing
  the blue (\textit{top}) and yellow (\textit{bottom}) WR bumps in
  Tol\,89-A. The blue bump is a blend of \textit{nebular} {\feiii}4658
  and \textit{stellar} {\nv}4620 {\niii}4640, {\ciii/\civ} \lam4650/58
  and {\heii}4686 -- indicating the presence of both early (WNE) and
  late-type (WNL) nitrogen rich Wolf--Rayet stars. The yellow bump is
  due to {\civ} emission from the strong stellar winds of WC4--5
  stars.}
\end{figure}

Regarding stellar features, we confirm the observations made by
\citeauthor{schaerer99}, detecting broad WR emissions in knot A at
{\lam}4640 (the blue bump) and {\lam}5808 (the yellow bump), see
Figs.~\ref{uves_a} and \ref{awrbumps}. The blue bump is a blend of
\textit{nebular} {\feiii}4658 and \textit{stellar} {\nv}4620,
{\niii}4640, {\ciii/\civ} \lam4650/58 and {\heii}4686, suggesting the
presence of both early (WNE) and late-type (WNL) nitrogen-rich WR
stars. The equivalent width of the blue bump is $\approx11${\AA}, with
\lam4686 contributing $\approx3${\AA}.  We also detect broad (FWHM
$\sim 80$ {\AA}) {\civ}5808 emission from WC stars with an equivalent
width of $\sim12$ {\AA}.

In region B only \textit{nebular} lines of {\feiii}4658, 4701,
{\ariv}4711/40 and {\hei}4713 are present in the region of the blue
bump, see Fig.~\ref{nebheii}b. No broad WR emission is seen.  Again,
we detect the presence of \textit{nebular} {\heii}4686.  The width of
{\heii}4686 emission is comparable to that of {\feiii} and {\ariv},
{\cf} $0.6\pm0.2$ and $0.7\pm0.1${\AA}, respectively. The origin of
this nebular {\heii} 4686 emission is discussed in
Section~\ref{discussion}.

\section{Properties of the knots and their ionizing clusters}
\label{tol89_properties}

In this section we derive the ages and extinction towards knots A and
B and their ionizing clusters A1--4 and B1 using \textit{stellar}
(Section~\ref{cluster_properties}) and \textit{nebular}
(Section~\ref{knot_properties}) diagnostics in the UV and optical
respectively. Cluster sizes for A1--4, B1, C1 and C2 are also derived
from the UV STIS image in Section~\ref{cluster_sizes}. Note that no
age or extinction estimates can be made for C1 and C2 since no
spectroscopic data is available for these clusters.

\subsection{UV derived cluster properties}
\label{cluster_properties}

\subsubsection{Sizes}\label{cluster_sizes}

Cluster sizes were determined from the STIS UV image using the
\textsc{ishape} routine of \citet{larsen}. The routine compares the
observed cluster profile with PSF-convolved analytical functions that
model the surface brightness of the clusters. The best-fitting
function, determined by the $\chi^{2}$ results, then gives the
effective radii $R_{\rm{eff}}$, or half light radius for each cluster.
A correction for the true effective radii for elliptical profiles
$R_{\rm{eff,ell}}$, as given by equation 11 of the \textsc{ishape}
user's guide, is made.

We choose to model our clusters using \citet{king} profiles for
different concentration parameters $c$ -- where $c$ is defined as the
ratio of the tidal radius $r_{\rm{t}}$ to the core radius $r_{\rm{c}}$
and takes values of 15, 30 and 100 -- and \textsc{moffat} profiles for
different power indices $\alpha$, where $\alpha$ equals 1.5 or 2.5
(\textsc{moffat15} or \textsc{moffat25}). Various fitting radii were
used, ranging from a minimum radius that is set equal to the
approximate size of the cluster to a maximum radius that is set to be
the furthest one can move from the cluster without background
contamination from a nearby source. The results of the models giving
the most internally consistent FWHM and best $\chi^{2}$ statistics are
shown in Table~\ref{fwhm}. All the clusters appear to be very compact,
with half light radii $\la3$ pc and sizes typical of SSCs
\citep{larsen04}.

\begin{table*}
\caption{Results of the best-fitting analytical profiles giving
the most internally consistent $\chi^{2}$ statistics and FWHM for clusters
A1--4, C1 and C2 seen in the UV STIS image. The FWHM and
minor-to-major axis ratios are averages of between 4--6
measurements over different fitting radii.}\label{fwhm}
\setlength{\tabcolsep}{2.6mm}
\begin{tabular}{|c|c|c|c|c|c|c|c|}
\hline\hline
\mrtwo{Cluster} &\mrtwo{Model}       
& FWHM & minor  & $R_{\rm{eff}}$ & $R_{\rm{eff,ell}}$ 
& $R_{\rm{eff,ell}}$   & $R_{\rm{eff,ell}}$ \\\cline{4-4}

& & (pix) & major & (pix) & (pix) & ($''$) & (pc) \\\hline
A1 & \textsc{moffat25}  & 2.4 & 0.79 & 1.63 & 1.46 & 0.036 & 2.6 \\
A2 & King15             & 2.4 & 0.73 & 1.69 & 1.46 & 0.036 & 2.6 \\ 
A3 & \textsc{moffat25}  & 2.8 & 0.75 & 1.92 & 1.68 & 0.041 & 3.0 \\
A4 & King15             & 1.0 & 0.76 & 1.05 & 0.92 & 0.023 & 1.6 \\ 
C1 & \textsc{moffat25}  & 2.5 & 0.51 & 1.67 & 1.26 & 0.031 & 2.2 \\
C2 & \textsc{moffat25}  & 2.0 & 0.71 & 1.37 & 1.17 & 0.029 & 2.1 \\
\hline
\end{tabular}
\end{table*} 

\subsubsection{Ages}\label{cluster_ages}

We adopt the method of \citet*{chandar04} for determining the ages of
our clusters by first correcting the STIS UV spectra for Galactic
foreground extinction. We use the \citet{seaton} Galactic law for an
{\ebvg}\,$=0.066$ mag \citep*{schlegal}. We then normalise the spectra
and compare them to the model data output of Starburst99\,v5.0
\citep{leitherer} scaled from a $10^{6}$ {\msun} instantaneous burst
with a 0.1--100 {\msun} \citet{kroupa01} IMF.

Based on our abundance determinations made in
Section~\ref{abundances}, an LMC metallicity was assumed.  UV O star
spectra shortward of $\lambda \leq$1600{\AA} are taken from template
stars in the Large and Small Magellanic Clouds \citep{leitherer01}.
All other empirical data are from spectral types at solar metallicity,
i.e.  the atlases of \citet*{robert93} for WR stars and
\citet*{demello00} for B stars.

\begin{figure} 
\includegraphics[scale=0.6,angle=0,clip=true]
{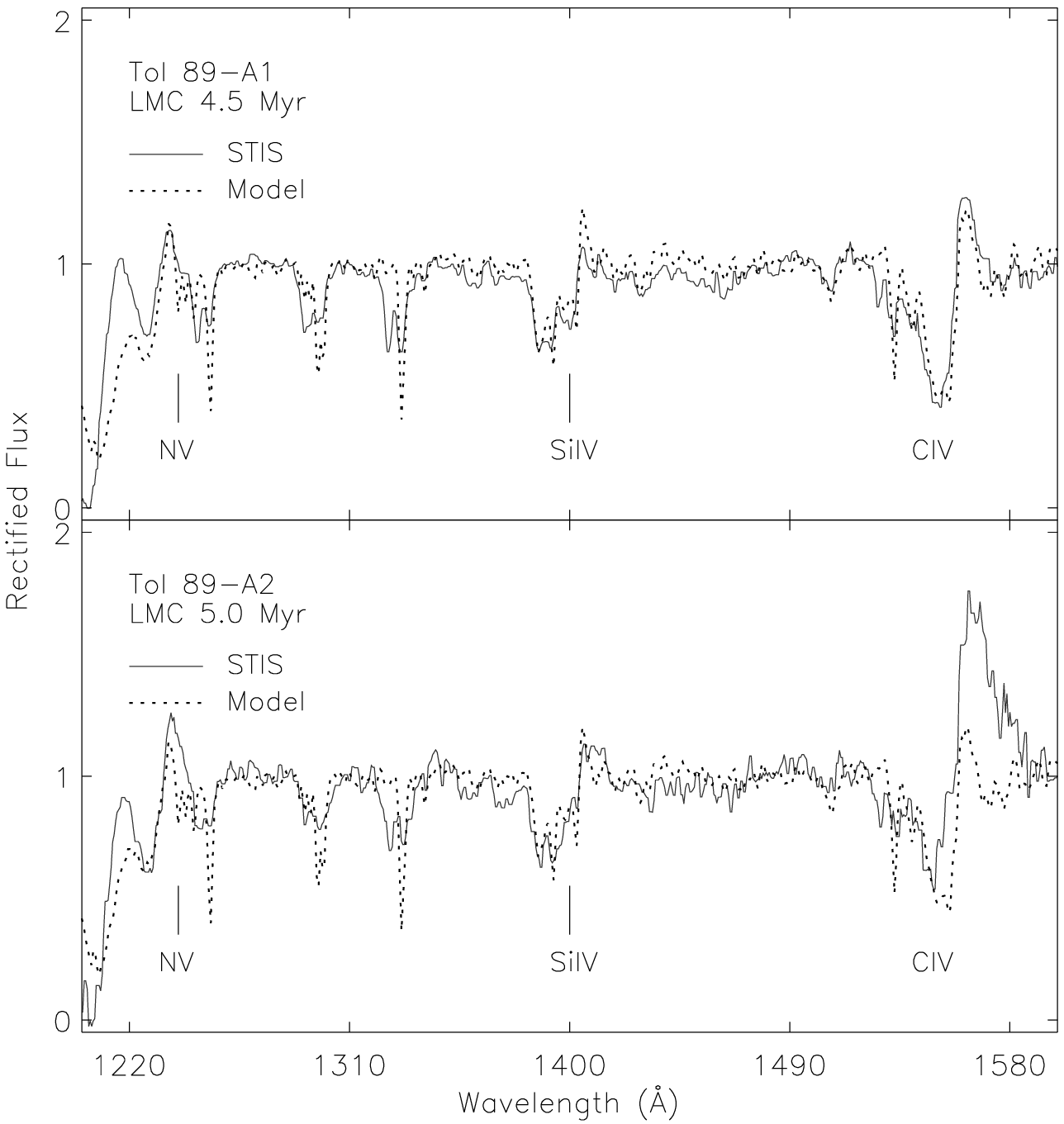}
\caption{\label{uv_norm} The rectified FUV spectra (\textit{solid
    black line}) of clusters A1 (\textit{top}) and A2
  (\textit{bottom}) shown with the best-fitting (age) Starburst99, LMC
  models (\textit{dotted line}). The STIS spectra have been smoothed
  for clarity and velocity corrected.}
\end{figure}

In Fig.~\ref{uv_norm} we show the best-fitting models for clusters A1
and A2. We can see in the case of A1 that we are able to reproduce
well the strengths of the {\siiv}1400 and {\civ}1550 O-star P-Cygni
profiles, enabling us to constrain the age to $4.5\pm0.5$ Myr. For A2
we determine a similar age of $5\pm1$ Myr. The uncertainty is larger
since we rely solely on the fit to {\siiv}1400.  This is due to the
fact that the Starburst99 models fail to reproduce the {\civ}1550
feature, whose strength implies a significant contribution from WC
stars in A2; unlike in A1 where the fit to {\civ}1550 is excellent.
The WR spectral synthesis in the UV is rudimentary because it relies
on solar metallicity spectra of a few WR stars. Until detailed WR
spectral synthesis is available for the UV wavelength range, it is
difficult to draw any quantitative conclusions on the WR content of
regions A1 and A2 from Starburst99 modelling alone. Our derived ages
for A1 and A2 are consistent with the $4\pm1$ Myr determined by
\citet{chandar04} for their extraction of Tol\,89-1 (= our A1 + A2).

Our age estimates for A3, A4 and B1 are somewhat more uncertain given
the poorer S/N of the spectra; we obtain approximate ages of 3.0--5.5
Myr for A3 and A4 and $<3$ Myr for B1 from the model fits. The age
estimate for A4 is especially uncertain, due to the weakness of
{\siiv}, with mid-WN stars contributing to the {\civ} profile, such
that Starburst99 modelling is inadequate. We suspect that A4 has an
age of $\sim$3 Myr, given its similarity to mid-WN stars observed in
young massive clusters (Section~\ref{est_nwr}).

Given the age of cluster A3 one would also expect WR stars to be
present. Whilst WC stars do not appear to be present from the observed
CIV 1550 profile, weak HeII 1640 from WN stars is not excluded given
the observed low S/N for this region. The presence of WN stars in A4
is discussed in Section~\ref{discussion}.

\subsubsection{Extinction}\label{cluster_extinction}

From the best-fitting age models we determine an estimate for the
internal reddening {\ebvi} towards each cluster by dereddening the
Galactic foreground extinction-corrected cluster spectra to match the
slope of the best fitting Starburst99 model. The fit is performed over
the wavelength range of 1240--1600 {\AA} and the models normalised to
match the continuum either side of {\civ}1550.  We adopted the stellar
LMC extinction law of \citet{howarth} to deredden our UV spectra in
favour of the \cite{calzetti} starburst obscuration law used by
\cite{chandar04}. The latter is more appropriate for unresolved star
forming galaxies than for effective point sources, such as is the case
for Tol\,89--A1 and A2. Similar conclusions were reached for
NGC\,3125--A1 by \cite{hc06}.

The following internal reddening values are obtained for A1 and A2
respectively: $0.09\pm0.02$ and $0.08\pm0.02$ mag. This is equivalent
to $\approx1.0\pm0.2$ and $0.9\pm0.2$ mags of internal extinction at
\lam1500 ($A_{1500}$) for the \citet{howarth} LMC reddening law.
Although our reddening estimates are in agreement with the value of
0.08~mag obtained by \citeauthor{chandar04}, the extinction at
\lam1550 obtained here is approximately twice that obtained using the
\citeauthor{calzetti} starburst obscuration law; for which $A_{1500}$
= 0.4 mag for {\ebvi}$=0.08$ mag. For regions A3, A4 and B1 we obtain
{\ebvi} of $0.10\pm0.03$, $0.08\pm0.03$ and $0.17\pm0.03$ mag, or
$A_{1500} \approx 1.1\pm0.3$, $0.9\pm0.3$ and $1.8\pm0.3$ mag
respectively. Again, using the \citeauthor{calzetti} law the
extinction at \lam1500 is approximately half that obtained using the
LMC law \citep{howarth}.

\subsection{Optically derived knot properties}
\label{knot_properties}

\subsubsection{Ages}\label{knot_ages}

Estimates for the ages of knots A and B were determined from the UVES
spectra by comparing the equivalent widths of nebular {\hb} emission
to Starburst99 predictions. We obtain equivalent widths of 55 and 295
{\AA} for A and B respectively.  For an instantaneous burst of LMC
metallicity and a 0.1--100 {\msun} Kroupa IMF, this corresponds to
ages of $\sim$4.5 and 2.5~Myr; assuming that all hydrogen ionising
photons are absorbed within the {\hii} region.  \citet*{schaerer}
determine an age of 4.5--5.0 Myr for Tol\,89 by comparing their
measured {\hb} equivalent widths with standard, SMC population
synthesis models.

The ages of the knots derived from the $EW(\hb)$ are in excellent
agreement with the cluster ages derived in Section~\ref{cluster_ages}.
The average cluster age in knot A (A1--4) is
$\approx4.5\pm1.0$ Myr, which is in excellent agreement with the 4.5
Myr determined for Tol\,89-A. An age $<3$ Myr is derived for B1, which
is also in good agreement with the $<2.5$ Myr derived for Tol\,89-B.

\subsubsection{Extinction}\label{knot_extinctions}

In the optical, the total extinction {\ebvt}\footnote{Here we define
  {\ebvt} to be the Galactic foreground contribution {\ebvg} (0.066
  mag; Schlegel et al.  1998) plus the internal contribution,
  {\ebvi}.}  is determined from the Balmer line decrement.  For both
the STIS and UVES data we use only the {\ha} and {\hb} line fluxes to
determine the extinction for the following reasons. In the STIS data,
the higher order Balmer lines of {\hg} and {\hd} are very weak and
thus are not reliable indicators of {\ebvt}.  In the UVES data we
observe little to no underlying stellar absorption. After correction
for galactic foreground extinction we determine an internal reddening
value for knot A which is consistent with there being zero reddening.
The low extinction towards knot A is surprising given it's young age
($\sim4.5$ Myr) and suggests that clusters very quickly disperse their
natal clouds within a few Myr.  For knot B, we obtain an internal
reddening value {\ebvi} of $\sim$ 0.24 from both the STIS and the VLT
spectra. We therefore adopt the following (total) reddening values for
knots A and B of $0.07\pm0.01$ and $0.29\pm0.03$ mag respectively.
Previous determinations of the average reddening over the Tol\,89
complex lie within the bounds of our estimates, {\cf} 0.12
(\citealt{terlevich91}: $C(\hb)=0.18$) and 0.20 \citep{durret}.

\section{Nebular properties of the knots}\label{nebular_properties}

Line fluxes and equivalent widths for nebular lines were determined
using the \textsc{elf} (emission line fitting) and \textsc{ew}
(equivalent width) routines within the \textsc{starlink} package
\textsc{dipso}. Fluxes were measured by Gaussian fitting, allowing for
line centres and widths to vary freely -- except when fitting doublets,
in which case widths and relative line centres were constrained. The
observed and intrinsic line fluxes normalised to {\hb}~=~100 are listed
in Table~\ref{fluxes}.

\begin{table}

\caption{Observed (F$_{\lambda}$) and intrinsic (I$_{\lambda}$) nebular 
line fluxes for knots A and B relative to {\hb}~=~100, and equivalent 
widths, W$_{\lambda}$, 
of WR emission lines seen in knot A. No WR emissions are detected
 in knot B.}\label{fluxes}
\begin{tabular}{ll@{\hspace{5mm}}r@{\hspace{5mm}}r@{\hspace{5mm}}c@{\hspace{5mm}}rr}\hline\hline
 & & \mctwo{A} & & \mctwo{B} \\\cline{3-4}\cline{6-7}
$\lambda_{0}$ & Ion & \mcone{F$_{\lambda}$} & 
\mcone{I$_{\lambda}$} & & \mcone{F$_{\lambda}$} & \mcone{I$_{\lambda}$}\\\hline
& & \mcfive{Nebular Emission Lines} \\\hline
3726 & {\oii}& 73.92 & 78.23 & &39.34 & 50.46 \\
3729 & {\oii}& 101.20 & 107.10 & &51.43 & 65.93 \\
3869 &{\neiii}& 26.33 & 27.70 & &27.95 & 34.95 \\
3967 &{\neiii}& 6.63 & 6.94 & &7.25 & 8.90 \\
4026 & {\hei}& 1.81 & 1.90 & &1.50 & 1.81 \\
4069 & {\sii}& 0.74 & 0.77 & &0.71 & 0.85 \\
4076 & {\sii}& ... & ... & &0.18 & 0.21 \\
4102 & {\hd} & 22.49 & 23.41 & &22.08 & 26.34 \\
4121 & {\hei}& ... & ... & &0.18 & 0.22 \\
4144 & {\hei}& ... & ... & &0.18 & 0.21 \\
4341 & {\hg} & 48.87 & 50.25 & &41.84 & 47.31 \\
4363 &{\oiii}& 2.28 & 2.34 & &2.60 & 2.92 \\
4471 & {\hei}& 3.22 & 3.29 & &3.40 & 3.73 \\
4658 &{\feiii}& 0.68 & 0.68 & &0.57 & 0.60 \\
4686 &{\heii}& ... & ... & &0.12 & 0.12 \\
4702 &{\feiii}& ... & ... & &0.10 & 0.10 \\
4711 &{\ariv}& 0.50 & 0.51 & &0.22 & 0.23 \\
4713 & {\hei}& ... & ... & &0.45 & 0.46 \\
4740 &{\ariv}& 0.32 & 0.33 & &0.24 & 0.25 \\
4959 &{\oiii}& 121.40 & 120.80 & &166.40 & 162.50 \\
5007 &{\oiii}& 363.20 & 360.30 & &498.50 & 481.50 \\
5016 & {\hei} & 2.18 & 2.16 & &2.06 & 1.99 \\
5048 & {\hei} & ... & ... & &0.11 & 0.10 \\
5518 &{\cliii}& 0.52 & 0.50 & &0.49 & 0.42 \\
5538 &{\cliii}& 0.41 & 0.39 & &0.31 & 0.26 \\
5755 & {\nii}& ... & ... & &0.46 & 0.38 \\
5876 & {\hei}& 11.60 & 11.07 & &15.07 & 12.24 \\
6300 & {\oi} & 1.85 & 1.74 & &1.71 & 1.30 \\
6312 &{\siii}& 1.42 & 1.33 & &2.02 & 1.54 \\
6364 & {\oi} & 0.22 & 0.21 & &0.53 & 0.40 \\
6548 & {\nii}& 4.53 & 4.23 & &5.42 & 3.98 \\
6563 & {\ha} & 290.20 & 270.40 & &390.50 & 286.30 \\
6583 & {\nii}& 14.91 & 13.88 & &17.26 & 12.62 \\
6678 & {\hei}& 3.09 & 2.87 & &4.80 & 3.47 \\
6716 & {\sii}& 13.29 & 12.33 & &16.31 & 11.71 \\
6731 & {\sii}& 9.97 & 9.24 & &12.80 & 9.18 \\
7065 & {\hei}& 2.44 & 2.24 & &4.29 & 2.95 \\
7136 &{\ariii}& 8.99 & 8.24 & &16.24 & 11.06 \\
7281 & {\hei}& 0.52 & 0.47 & &0.91 & 0.61 \\
7319 & {\oii}& 1.56 & 1.42 & &0.79 & 0.52 \\
7331 & {\oii}& 0.84 & 0.76 & &1.19 & 0.79 \\
9069 &{\siii}& 27.70 & 24.35 & &56.23 & 31.90 \\
9531 &{\siii}& 97.78 & 85.32 & &199.60 & 109.70 \\
 & & & & & &\\
\multicolumn{2}{l}{$\log$ \hb} & --13.32 & --13.23& 
& --12.86 & --12.43 \\
\multicolumn{2}{l}{\ebvt} & \mctwo{0.07} & & \mctwo{0.29} \\ \hline

& & \mcfive{Broad WR Emission Lines in Region A} \\\hline

& & \multicolumn{2}{l}{I$_{\lambda}$ ($10^{-15}${\ergss})} 
& & \mctwo{W$_{\lambda}$ ({\AA})} \\

\multicolumn{2}{l}{\nv $\lambda$4606, 4619} 
& \mctwo{3.6$\pm$0.6} & & \mctwo{2.2$\pm$0.4} \\
\multicolumn{2}{l}{\niii $\lambda$4634--41} 
& \mctwo{2.1$\pm$0.6} & & \mctwo{1.2$\pm$0.4} \\
\multicolumn{2}{l}{\ciii/\civ $\lambda$4650/58} 
& \mctwo{3.4$\pm$0.4} & & \mctwo{2.1$\pm$0.2}\\
\multicolumn{2}{l}{\heii $\lambda$4686} 
& \mctwo{3.9$\pm$0.4} & & \mctwo{2.6$\pm$0.3} \\
\multicolumn{2}{l}{\civ $\lambda$5808} 
& \mctwo{5.9$\pm$1.4} & & \mctwo{11.8$\pm$0.7} \\\hline

\end{tabular}

\end{table}   

We find that the strongest nebular emission lines are composed of a
narrow and a broad component, both centred at similar velocities. In
addition for knot A, we detect line splitting in the narrow component
with a separation between the blue (V1) and red (V2) components of
$48\pm1$ {\kms}. The FWHMs of the broad and narrow components are for
knot A: $111\pm2$, $32\pm2$ (blue component), $41\pm2$ {\kms} (red
component), and $71\pm2$, $27\pm2$ {\kms} for knot B. The multiple
components are seen in the following emission lines: {\ha}, {\hb},
{\oii} \lam\lam\,3726/29, \lam\lam\,7319/30, {\oiii}\lam\lam\,4959,
5007, {\nii}\lam\lam\,6548/84, and {\sii}\lam\lam\,6717/31. In lines
with poorer S/N no broad or velocity-split components are detected.

\begin{figure}
\includegraphics[scale=0.6,angle=0,clip=true]
{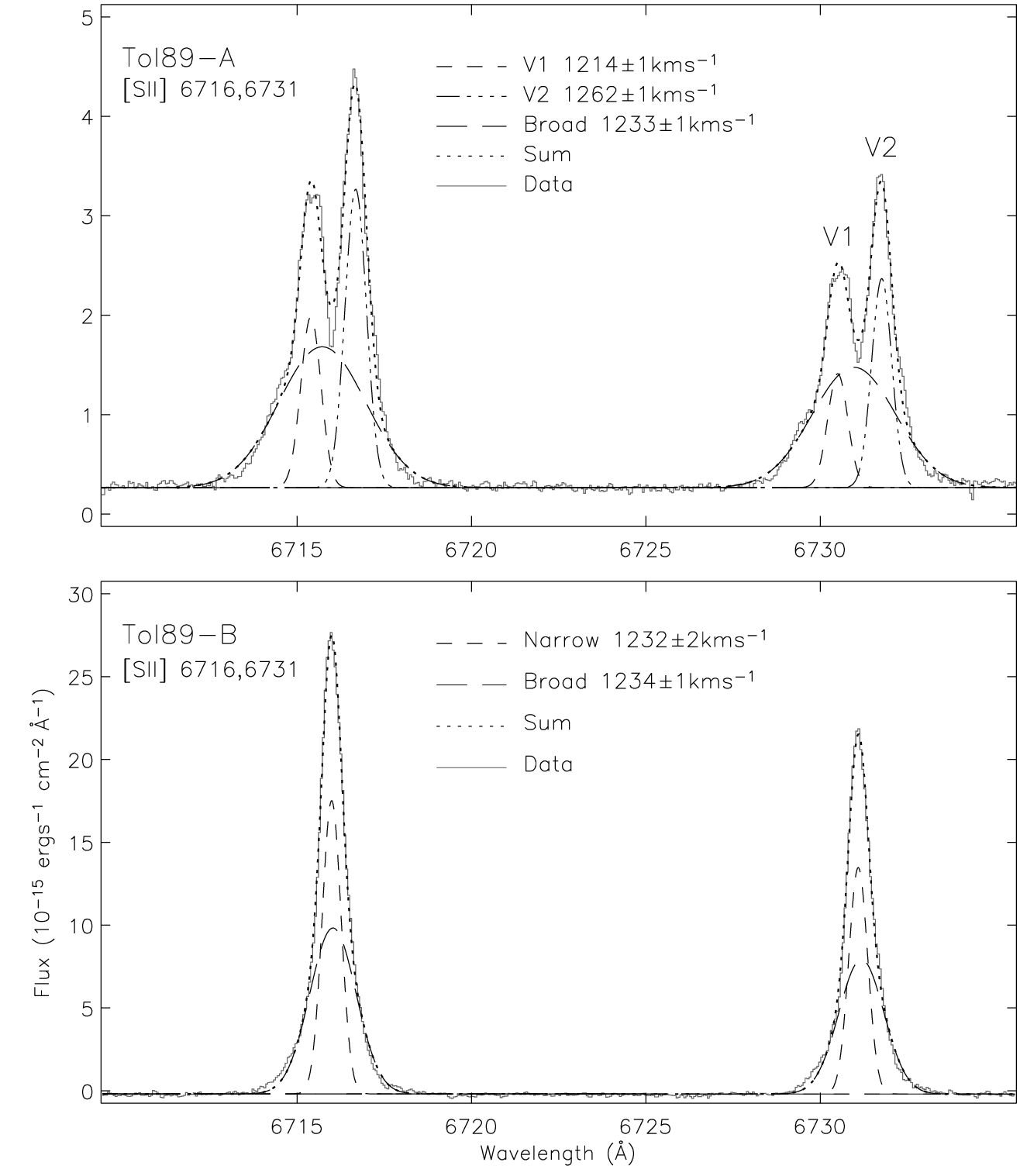}
\caption{\label{line_profile} UVES spectra of knots A
  (\textit{top}) and B (\textit{bottom}) in the spectral range of the
  {\sii} 6716, 6731 doublet showing the multiple Gaussian component
  fits to the line profiles. The broad (\textit{long--dash}), narrow
  (\textit{short--dash}) and summed components (\textit{dotted
    line}) are plotted. In region A, the two narrow velocity
  components, V1 (\textit{short-dashed}) and V2
  (\textit{dash--dot--dot--dot line}) can be seen. The central
  velocities of the Gaussian fits are given in each plot.}
\end{figure}

In Fig.~\ref{line_profile} we show, as an example, the {\sii}
\lam\lam\,6716/31 emission seen in A ({\textit{top}}) and B
({\textit{bottom}}) with the best-fitting Gaussians superimposed. The
broad components account for a significant fraction of the total line
flux (31 and 50 per cent for A and B respectively). The broad
components are symmetrical with respect to the mean velocities of the
narrow components and have the same central velocities for both A and
B, suggesting a common origin.  The only apparent difference between
the broad components in the A and B spectra is in the line width (111
vs. 71 {\kms}).

Mean heliocentric radial velocities for A and B were determined by averaging
the velocities of all the Gaussian components (broad and narrow) fit
to the emission lines in each region and are found to be $1241\pm2$
and $1234\pm2$ {\kms} respectively. These results are consistent with
the values reported in the literature: $1232\pm51$ \citep{durret} and
$1226\pm11$ {\kms} \citep{schaerer}. 

Overall, we find that knots A and B have similar velocities, differing
by 7 {\kms}, and thus are part of the same star-forming complex. For
A, we see velocity splitting, suggesting that the winds from the
clusters have swept up the surrounding interstellar gas into a shell
which is currently expanding at 24 {\kms}. For B, we see no line
splitting which is in accord with its younger age. We also detect a
broad component which appears to be common to both regions A and B.
The maximum width of these features corresponding to the full width at
zero intensity (FWZI) is 1140 and 940 {\kms} for A and B; this
suggests that the two knots contain gas with velocities up to $\sim$
450--600 {\kms}.  We consider the origin of this high velocity
component in Sect.~\ref{discussion}.

\subsection{Electron densities and temperatures}\label{nete}

Electron densities {\ne}, and temperatures {\te}, were determined
using the diagnostic line ratios listed in Table~\ref{lineratios}.
The individual \textit{broad} and \textit{narrow} velocity components
were summed in our calculation of {\ne} and {\te} due to the fact that
no broad or velocity shifted components were detected for the weaker,
poorer S/N, temperature-sensitive diagnostic lines (\lam\lam4069,
4076, 4363 and 5755). We note that the ratios of the line fluxes for
the broad and narrow components in {\sii} are identical (see
Fig.~\ref{line_profile}) suggesting that they have similar electron
densities.

Electron densities were calculated assuming a constant {\te} of
10\,000 K. Since {\oii} and {\sii} give consistent {\ne} (see
Table~\ref{lineratios}) we take the average of the two to be
representative of the densities in Tol\,89-A and B. We find {\ne} of
$\approx90\pm40$ and $\approx150\pm40$ {\cmq} respectively.  {\ariv}
and {\cliii} probe denser regions of gas and thus give higher {\ne}
estimates. Refined estimates for the electron temperatures were then
determined using the average {\ne} determined from O$^+$ and S$^+$
above. Using the most consistent {\te} between the different
diagnostic line ratios, we derive the following `average' {\te} of
$\approx10\,000\pm300$ K for A; neglecting $T(S^+)$ due to its
large uncertainty.

For region B we derive a value of $9\,800\pm300$ K. We exclude the
large {\te} derived from N$^+$ and O$^+$ because in moderate to
high-excitation {\hii} regions the {\nii}5755 and {\oii}7330 lines can
be excited by the recombination of N$^{2+}$ and O$^{2+}$ in the
higher-excitations zones \citep{rubin86,liu00}, thus leading to higher
electron temperatures (see Table~\ref{lineratios}).

\begin{table}
\caption{Derived physical properties and abundances from the
UVES spectra of Tol\,89 for knots A and B. A comparison is made with the 
abundances derived for the LMC \citep{russell90} and 30\,Dor
\citep{peimbert03}.}\label{lineratios}
\setlength{\tabcolsep}{1mm}
\begin{tabular}{|l|c|c|c|c|}
\hline\hline
Diagnostic & \mctwo{Knot} & &\\
Line Ratio & A     & B & LMC$^a$ & 30\,Dor$^b$ \\\hline
Density &  \mctwo{\ne(\cmq)} & &\\
{\oii}\lam3726/\lam3729 & 90$^{+40}_{-50}$ & $140\pm30$ & &\\
{\sii}\lam6731/\lam6716 & $100\pm30$ & $150\pm50$ & &\\
{\ariv}\lam4740/\lam4711  & ...   & 4300$^{+2000}_{-1800}$ & &\\
{\cliii}\lam5537/\lam5517 & 510--1450  & ...   & &\\
 & & & &\\
Temperature & \mctwo{\te(K)} & &\\
{\nii}\lam5755/\lam6584   & ...   & 14800$\pm$400 & &\\
{\oii}\lam7330/\lam3726   & 9900$\pm$200  & 13100$\pm$400 & &\\
{\sii}\lam4068/\lam6717   & 9400$\pm$900  & 10000$\pm$500 & &\\
{\oiii}\lam4363/\lam5007  & 10000$\pm$500 & 9800$\pm$100 & & \\
{\siii}\lam6312/\lam9069  & 10000$\pm$200 & 9500$\pm$100 & & \\
 & & & & \\
Average                   &       &       & &\\
{\ne}(\cmq) & $90\pm40$ & $150\pm40$ & &\\
{\te}(K) & 10000$\pm$300 & 9800$\pm$300 & &\\
 & & & & \\
Abundances   & & & & \\
O$^{+}$/H$^{+}$(x10$^{4}$) & 0.68$\pm$0.09  & 0.49$\pm$0.06 & &\\
O$^{2+}$/H$^{+}$(x10$^{4}$) & 1.21$\pm$0.17  & 1.91$\pm$0.17 & &\\
O/H(x10$^{4}$) & 1.88$\pm$0.36  & 2.40$\pm$0.35 & &\\
12+log(O/H)  & 8.27$^{+0.08}_{-0.09}$ & 8.38$^{+0.06}_{-0.07}$ & 8.37 & $8.50\pm0.02$\\
 & & & &\\
S$^{+}$/H$^{+}$(x10$^{6}$) & 0.46$\pm$0.03  & 0.51$\pm$0.04 & &\\
S$^{2+}$/H$^{+}$(x10$^{6}$) & 2.94$\pm$0.15 & 4.23$\pm$0.23 & &\\
ICF(S) & 1.28  & 1.57 & &\\
S/H(x10$^{6}$) & 4.37$\pm$0.36 & 7.47$\pm$0.66 & &\\
12+log(S/H) & 6.74$^{+0.03}_{-0.04}$ & 6.87$\pm$0.04 & 6.87 & $6.99\pm0.10$\\
 & & & &\\
N$^{+}$/H$^{+}$(x10$^{6}$) & 1.97$\pm$0.13 & 2.10$\pm$0.15 & &\\
ICF(N) & 2.78  & 4.92 & &\\
N/H(x10$^{6}$) & 5.47$\pm$0.37 & 10.30$\pm$0.73 & &\\
12+log(N/H) & 6.74$\pm$0.03  & 7.01$\pm0.03$ & 7.07 & $7.21\pm0.08$ \\
 & & & &\\
Ar$^{2+}$/H$^{+}$(x10$^{7}$) & 7.19$\pm$0.36  & 6.67$\pm$0.51  & &\\
Ar$^{3+}$/H$^{+}$(x10$^{7}$) & 0.63$\pm$0.11  & 0.56$\pm$0.07  & &\\
ICF(Ar) & 1.14  & 1.04  & &\\
Ar/H(x10$^{7}$) & 8.91$\pm$1.46  & 7.52$\pm$1.02 & &\\
12+log(Ar/H) & 5.95$^{+0.07}_{-0.09}$ & 5.88$^{+0.06}_{-0.07}$ & 6.07 & $6.26\pm0.10$ \\
\hline
\end{tabular}

\medskip
$^{a}$ Values taken from \cite{russell90}.\\
$^{b}$ Values taken from \cite{peimbert03} adopting $t^2=0.033$.

\end{table} 

\subsection{Abundances}\label{abundances}

Ionic abundances were calculated using the average {\ne} and {\te}
determined in Section~\ref{nete} (see Table~\ref{lineratios}). To
determine total abundances we adopt the following expressions given by
Eqn.~\ref{abundratio}; where the ionisation correction factors (ICF)
for N, S and Ar are as defined in equations 15, 18 and 19,
respectively, of \citet*{izotov94}.

\begin{displaymath}
\frac{O}{H} = \frac{ O^+ + O^{2+} }{H^+} \nonumber\\
\end{displaymath}

\begin{displaymath}
\frac{S}{H} = ICF(S)\frac{ S^+ + S^{2+} }{H^+} \nonumber\\
\end{displaymath}

\begin{displaymath}
\frac{N}{H} = ICF(N)\frac{ N^+ }{H^+} \nonumber\\
\end{displaymath}

\begin{eqnarray}\label{abundratio}
\frac{Ar}{H} = 
ICF(Ar)\frac{ Ar^{2+} + Ar^{3+} }{H^+} \nonumber\\
\end{eqnarray}

We derive 12+log(O/H) for knots A and B respectively of 8.27 and 8.38
which are in good agreement with the value of 8.32 derived by
\citet{schaerer}; while \citet{durret} derive a much lower oxygen
content of 8.03. Oxygen and sulphur abundances are very similar to
those of the LMC \citep{russell90}, such that hereafter we assume an
LMC metallicity for Tol\,89, although nitrogen and argon abundances
are somewhat lower. With respect to 30~Dor \citep{peimbert03}, knots A
and B are moderately depleted by 0.25--0.5~dex.

\section{Massive Star Population}\label{mpop}

In the following section we examine the massive star population in
Tol\,89, applying both empirical and synthesis techniques in the
optical and the UV.

\begin{figure*}
\includegraphics[scale=0.7,angle=0,clip=true]
{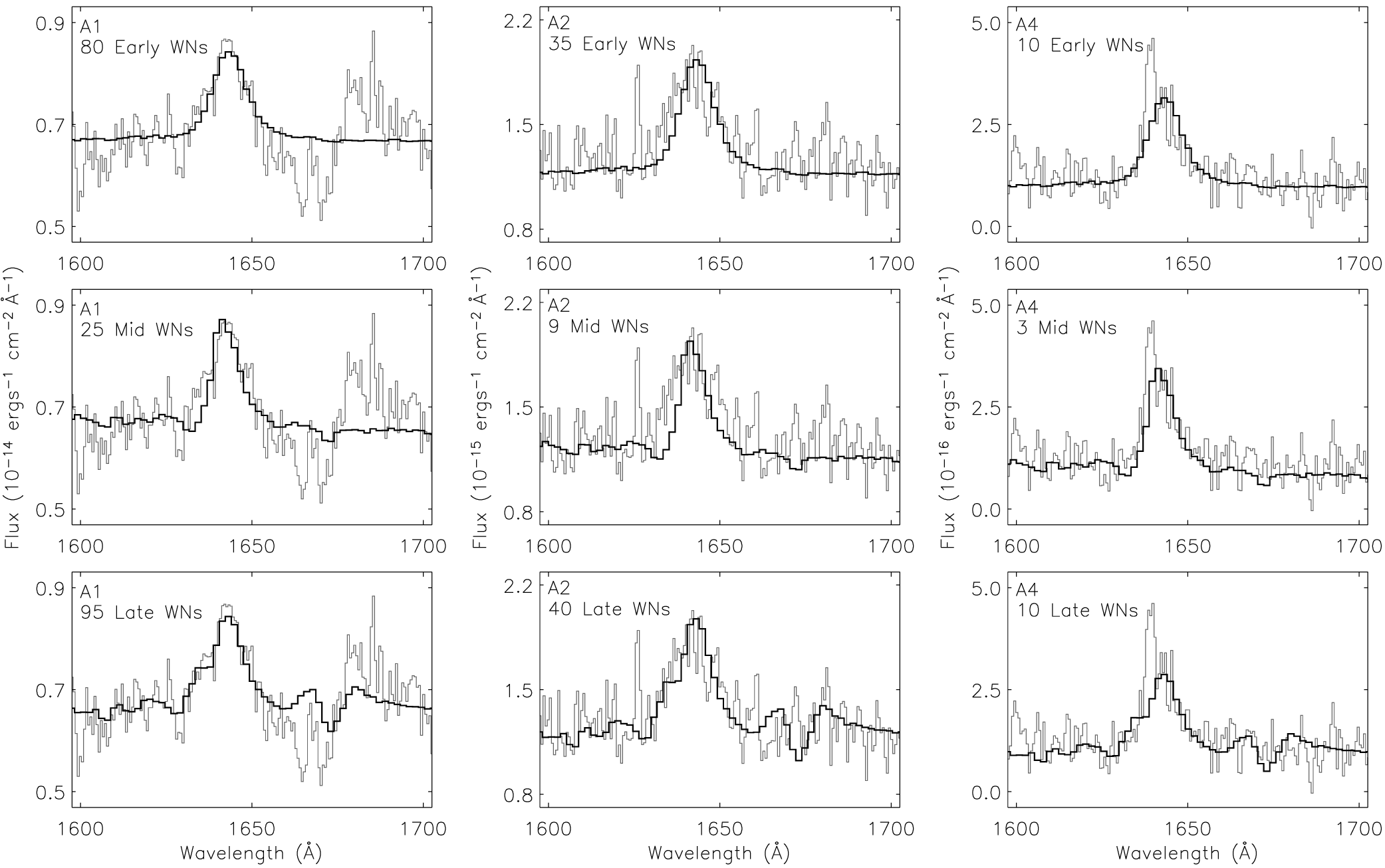}
\caption{\label{heii1640} Plots showing the fits to {\heii}1640 for 
  clusters A1 (\textit{left panel}), A2 (\textit{centre panel}) and A4
  (\textit{right panel}) using the generic LMC UV spectra presented in
  \citet{ch06} of early (\textit{top}), mid (\textit{middle}) and
  late-type (\textit{bottom}) WNs.}
\end{figure*}

\subsection{Empirical technique: line luminosities}\label{empirical}

\subsubsection{Determining the dominant WR subtypes}\label{wr_subtype}

Empirically, the WR population is estimated from average line
luminosities of WR stars and is thus dependent on the dominant
subtypes assumed to be present \citep{sv98}.

For WN stars we use the line width of {\heii}1640 in the UV as the
main discriminator of subtype, since it can be attributed almost
entirely to WN stars; in optical low resolution data, \lam4686 is
blended with nebular emission lines ({\feiii} and {\ariv}) or with
other WN \textit{and} WC line components ({\eg} {\niii}4634--41 and
{\ciii/\civ} \lam4650/58). Using the generic LMC UV spectra presented
in \citeauthor{ch06} (2006; hereafter CH06) of early, mid and late WNs
we reproduce the line morphology of {\heii}1640 in A1, A2 and A4 (see
Section~\ref{est_nwr}).

In Fig.~\ref{heii1640} we show the fits obtained to {\heii}1640. The
apparently good fit from the late-type templates is due to the poorer
resolution of the template spectra in which the apparent {\heii}1640
line width is due to blending with Fe\,{\sc iv}\,1632. A dominant
late-type population can be excluded since \lam1632, if present, would
be separated from \lam1640 at the resolution of the STIS UV spectra
(3.1 {\AA}). Although a dominant population of mid-type WNs cannot be
ruled out entirely, the width of the template line profile is narrower
than observed in either A1 or A2. We therefore conclude that the
dominant WN population in A1 and A2 is early-type from the fit to
{\heii}1640\footnote{A significant contribution to the {\heii}1640
  emission line flux in A2 comes from WC stars and thus the width of
  \lam1640 is less reliable as a discriminator of subtype in this
  case.}. For cluster A4 an excellent fit to the {\heii} 1640 line
profile is obtained using the mid-type templates (see \textit{middle
  right panel} of Fig.~\ref{heii1640}) and we therefore conclude that
WN5--6 stars dominate in A4.

In contrast, for WC stars we use the optical wavelength regime to
determine the main subtype because {\civ}5808 can be attributed solely
to WC stars, whilst {\civ}1550 suffers from contamination by O and WN
stars.  From the presence of \lam5808, and the absence of {\ciii}5696
emission, we conclude that WC4--5 stars are the dominant subtype.
This is supported by comparisons of our measurement of the FWHM of the
{\lam}5808 feature ($\sim80\pm10$ {\AA}) with those made by
\citet{crowther98}. We also infer that WC stars are solely present in
cluster A2 from the strength of {\civ}1550 relative to {\heii}1640 in
the ultraviolet.  \textit{In summary, we assume that early WN (WN2--4)
  and WC (WC4) stars dominate the WR populations of A1 and A2, with
  WC4 stars absent in A1, while in A4 we assume that mid WN(WN5--6)
  stars dominate.}

\subsubsection{Estimating the number of WR stars}\label{est_nwr}

\paragraph*{UV:}

\begin{figure}
\includegraphics[scale=0.63,angle=0,clip=true]
{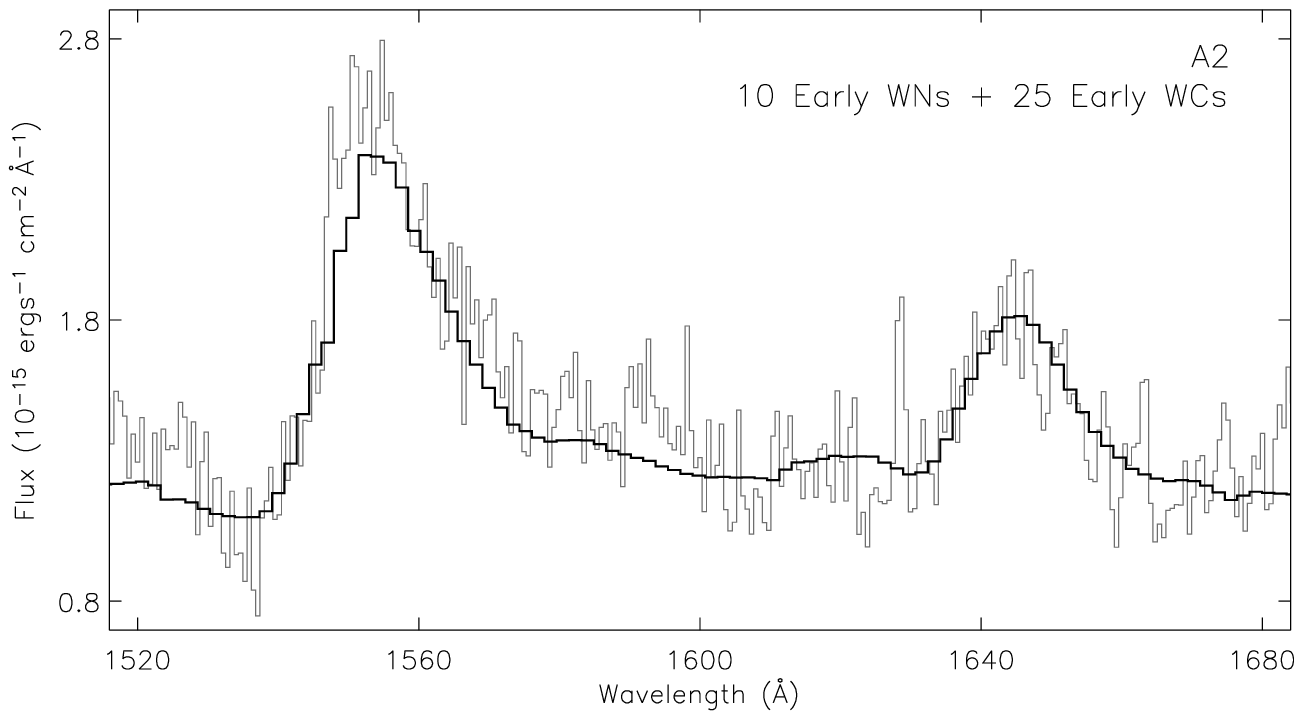}
\caption{\label{civ1550} Plot showing the fit to {\civ}1550 for 
  cluster A2 using the generic LMC UV spectra presented in
  \citet{ch06} for a mixed population of early-type WN2--4 and WC4
  stars.}
\end{figure}

We estimate the WR populations of A1, A2 and A4 by scaling
the generic LMC UV spectra presented in CH06 to match the intrinsic
line morphologies of {\heii}1640 and {\civ}1550, as shown in
Figs.~\ref{heii1640} and \ref{civ1550}. A comparison of this technique
versus simple line flux measurements is given in CH06.

For cluster A1 we estimate an early WN content of 80 stars, with WC
stars absent.  For A2, we estimate the WR content by simultaneously
fitting {\civ}1550 and {\heii}1640 until the strength of {\heii}1640
is reproduced, although we anticipate under-predicting {\civ}1550
emission since O stars will also contribute to this line (recall
Fig.~\ref{uv_norm}). We estimate 10 early WN stars plus $\sim$25 early
WC stars in A2. Remarkably, in cluster A4 we have been able to detect
just three mid WN stars despite the large distance of 14.7~Mpc to
Tol\,89, which is testament to the dominance of WR stars in integrated
(cluster/galaxy) spectra at UV wavelengths. To permit direct
comparisons with the WR populations inferred from the optical UVES
spectra of knot A, we sum the individual WR numbers in A1 and A2, as
indicated in Table~\ref{wrtemp}; we do not include the three WN5--6
stars derived for cluster A4 since these will not contribute
significantly to the optical blue bump which we attribute entirely to
WN2--4 stars.

Previously, \cite{chandar04} derive a value of $95\pm68$ late-type WN
stars based on the their de-reddened {\heii}1640 line flux for
Tol\,89--1 (which includes our clusters A1 and A2) and on the average
{\heii}1640 line luminosity of a WNL star taken from SV98
(L$_{1640}=1.2\times10^{37}$ {\ergss}). From the line luminosity of
{\lam}1640 given in their table 3, this equates to $\sim130$ early WN
stars using the CH06 LMC line luminosity for WN2--4 stars, a factor of
1.3 times larger than our estimate of $\sim100$. We have measured the
observed {\heii} line flux from the spectrum of \citet{chandar04},
kindly made available to us by the author, and have measured an
observed flux that is in agreement with the sum of the observed fluxes
in A1 and A2. Using the LMC template spectra of CH06, we derive an
early WN content of $\sim90$ in agreement with the results obtained
for the sum of A1 and A2. Thus, we attribute this difference in WN
numbers ({\cf} $\sim100$ and 130) to the different choice of
extinction laws and techniques adopted.

\paragraph*{Optical:}

Applying the same techniques in the optical, we estimate the number of
WR stars from fits to {\heii}4686 (WN+WC) and {\civ}5808 (WC) seen in
knot A. Recall, no broad WR emission features are detected in knot B.
Since WC stars are known to contribute to the strength of \lam4686, we
begin by first estimating the number of WC stars responsible for the
yellow bump (\lam5808), so that their contribution to \lam4686 can
first be taken into account.

\begin{figure}
\includegraphics[scale=0.73,angle=0,clip=true]
{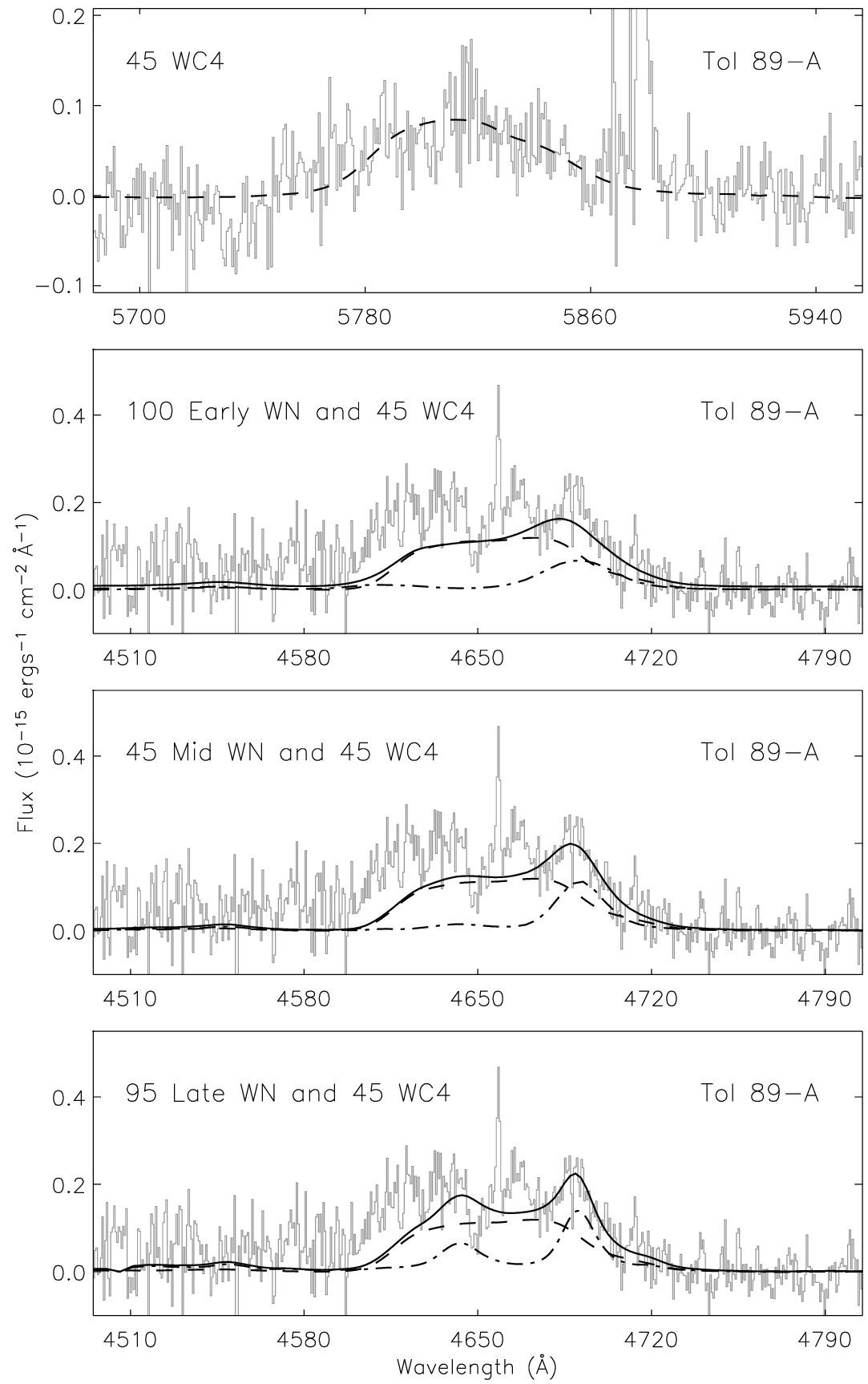}
\caption{\label{empsynth_fig} Plots showing the results of our line profile
  fitting. \textit{Top:} Fit to the {\civ}5808 feature using 45 LMC
  WC4 stars (\textit{dashed line}). \textit{2nd:} Plot showing the
  individual contributions from 100 early-type WNs (\textit{dot-dashed
    line}) and 45 early type WCs (\textit{dashed line}) to the blue
  bump. The sum of the individual components is shown as the
  \textit{solid} line. \textit{3rd:} As for \textit{panel 2} but
  showing the individual contributions from 45 mid-type WNs
  (\textit{dot-dashed line}). \textit{Bottom:} As for \textit{panel 2}
  but showing the individual contributions from 95 late-type WNs
  (\textit{dot-dashed line}). The UVES spectra have been binned for
  clarity, continuum subtracted and velocity and extinction
  corrected.}
\end{figure}

The results of our optical fitting procedure are shown in
Fig.~\ref{empsynth_fig}. From the fit to {\civ5808} (\textit{top
  panel, dashed line}) we estimate a population of 45 early WC stars
({\cf} 25 from the UV). Although the early-type empirical fit does not
give the best match to \lam4686, a dominant population of mid-type WNs
can be ruled out from the absence of {\niv}4058, while the UV excludes
a late-type dominant population from the absence of Fe\,{\sc iv}\,1632
emission.  Taking into account the 45 early WC stars determined from
{\civ}5808, we derive a population of 100 early-type WN stars, which
is in good agreement with the results obtained from the UV ({\cf} 90
WNE for A1+A2), as shown in Table~\ref{wrtemp}.

\begin{table}
\caption{WR content of Tol\,89-A derived using the average LMC template 
WR spectra and line luminosities (erg s$^{-1}$) of 
\citet{ch06}.}\label{wrtemp}
\setlength{\tabcolsep}{1.3mm}
\begin{tabular}{lllcccr}
\hline\hline
 &Diagnostic & Line       & A1 & A2 & A4 & \mcone{A} \\
 &Line       & Luminosity &    &    &    &           \\\hline   
& \mcsix{\textit{Optical}} \\ 
N(WN2--4) & $\lambda$4686 &8.4$\times 10^{35}$ &... &...  &...  & 100 \\  
N(WC4)    & $\lambda$5808 &3.3$\times 10^{36}$ &... &...  &...  & 45 \\
N(WR)     &               &                    &... &...  &...  & 145 \\
N(WC)/N(WN)&              &                    &... &...  &...  & 0.5 \\
& & & & & \\
&\mcsix{\textit{UV}}\\
N(WN2--4) & $\lambda$1640 &8.4$\times 10^{36}$ & 80 & 10  &...  & 90  \\
N(WN5--6) & $\lambda$1640 &1.7$\times 10^{37}$ &... &...  & 3   & 3  \\
N(WC4)    & $\lambda$1550 &2.0$\times 10^{37}$ &... & 25  &...  & 25  \\
N(WR)     &               &                    & 80 & 35  & 3   & 118 \\
N(WC)/N(WN)&              &                    &... & 2.5 &...  & 0.3 \\
\hline
\end{tabular}
\end{table}

In summary, our UV and optical WR diagnostics give consistent
populations of WN stars, plus reasonable agreement for WC stars. For
Tol\,89-A (which encompasses A1--4), our derived WR content of
$\sim95$ WNE and $\sim35$ WCE stars indicates N(WC)/N(WN) $\sim0.4$
\citep[{\cf} $\sim$0.6;][]{schaerer99}.

\subsubsection{Estimating the number of O stars}\label{est_no}

O-star numbers were derived following the methods outlined in
\citealt{sv98} (hereafter SV98) using the observed line luminosity of
{\hb}, which gives $Q_0^{\rm{obs}}$ -- the total number of hydrogen
ionising photons.  $Q_0^{\rm{obs}}$ is then related to the total
number of ionising O stars N(O) for a given subtype (typically
O7V) using Eqn.~\ref{opop2}; where the ionising contribution from a
population of WR stars is taken into account:

\begin{equation}\label{opop2}
N(O) = \frac{Q_0^{\rm{obs}} - 
N(WN)Q_0^{\rm{WN}} - 
N(WC)Q_0^{\rm{WC}}}
{\eta_0(t)Q_0^{\rm{O7V}}},
\end{equation}
and where $Q_0^{\rm{O7V}}$ is the Lyman continuum flux of an
individual O7V star and $\eta_0(t)$ represents the IMF-averaged
ionising Lyman continuum luminosity of a ZAMS population normalised to
the output of one equivalent O7V star \citep{vacca94}. N(WN)
and N(WC) are the number of WN and WC stars respectively.

From the line luminosities of {\hb} given in Table~\ref{olf} we derive
the following $Q_0^{\rm{obs}}$ values of $\sim320$ and
$\sim2000\times10^{49}$\,s$^{-1}$ for knots A and B, respectively.
The sum of these is $\sim3$ times that obtained by \cite{schaerer99}
who derived a $Q_0^{\rm{obs}}$ value of $708\times10^{49}$\,$s^{-1}$
for the Tol\,89 complex. This is likely due to the different choice
in slit width (1\farcs6) and PA (39{\degr}) ({\cf} Table~\ref{obs}).

Values of $\eta_0(t)$ are taken from the instantaneous burst models of
SV98 (see their fig. 21) for the ages given by the equivalent width of
{\hb} obtained in Section~\ref{knot_ages}. We adopt $\eta_0(t)$
values of 0.25 and 0.9 for knots A and B respectively (see
Table~\ref{olf}). From a calibration of line blanketed Galactic O
star models, $\log Q_0$(O7V) = 48.75 $s^{-1}$ \citep{martins05}.
Recent studies of Magellanic Cloud O stars indicate $\sim$2--4~kK
higher temperatures than their Galactic counterparts
\citep{massey05,heap06}. Consequently, we adopt $\log Q_0$(O7V) = 48.9
$s^{-1}$ at LMC metallicity \citep{hc06}. For typical LMC early WN and
early WC star populations we adopt $\log Q_0$ values of 49.0
\citep{crowther96} and 49.4 \citep{crowther02}, respectively.

For knot A we derive N(O)$\sim$ 690 from nebular {\hb} emission,
assuming a contribution from 95 WNE and 35 WCE stars giving N(WR)/N(O)
$\sim0.2$ (see Table~\ref{olf}). For region B we derive a content of
$\sim2800$ O stars from $Q_0^{\rm{obs}} \sim
2000\times10^{49}$\,$s^{-1}$ and $\eta_0(t)$ determined from the SV98
models (see their fig. 21).  This is about a factor two smaller than
the value obtained by \citeauthor{johnson03} using radio diagnostics
($\sim\,3800\,\times\,10^{49}s^{-1}$, given a distance of 14.7\,Mpc
for $H_{0}$\,=\,75\,\kms\mpc; \citeauthor{schaerer99}). We expect that
the discrepancy between our derived value of $Q_0^{\rm{obs}}$ with
that of \citeauthor{johnson03} is due to the fact that radio
observations probe deeper into the star-forming region, thus detecting
massive stars that are otherwise obscured at optical wavelengths.

\begin{table}
\caption{Nebular derived O star content for Tol\,89-A and B based on the line 
luminosity and equivalent width of {\hb}: L(\hb) and W({\hb}) 
respectively. The {\hb} line fluxes (F$_{\lambda}$ and I$_{\lambda}$) 
are in units of ergs\,s$^{-1}$\,cm$^{-2}$. Line luminosities are derived 
based on D\,$=14.7$ Mpc and are in units of {\ergss}. 
The WR numbers for Tol\,89-A are averages
of UV and optical empirical calibrations from 
Table~\ref{wrtemp}, whilst WR stars are not seen in 
Tol\,89-B.}
\label{olf}
\begin{tabular}{lll}
\hline\hline
Region & A & B \\\hline
F(\hb) & $4.7\times10^{-14}$ & $1.4\times10^{-13}$ \\    
I(\hb) & $5.9\times10^{-14}$ & $3.7\times10^{-13}$ \\   
L(\hb) & $1.5\times10^{39}$  & $9.5\times10^{39}$ \\
$W(\hb)$[\AA] & 55 & 295 \\ 
Age (Myr) & 4.5 & 2.5 \\    
$Q_0^{\rm{obs}}$ & $3.2\times10^{51}$ & $2.0\times10^{52}$ \\
$Q_0^{\rm{O7V}}$ & $8.0\times10^{48}$ & $8.0\times10^{48}$ \\
$Q_0^{\rm{WN}}$  & $1.0\times10^{49}$ &\\
$Q_0^{\rm{WC}}$  & $2.5\times10^{49}$ &\\
$\eta_0(t)$ & 0.25 & 0.9 \\  
$N_{\rm{WN}}$ & 95 &... \\ 
$N_{\rm{WC}}$ & 35 &... \\ 
$N_{\rm{O}}$  & 685 & 2780 \\ 
WR/O & 0.2 &...\\
\hline
\end{tabular}
\end{table}

\subsection{Synthesis technique: Starburst99 modelling}\label{sb99}

In this section we compare the observed optical UVES spectrum of
Tol\,89-A and B to model predictions computed using the evolutionary
synthesis code Starburst99\,v5.0 \citep{leitherer}. The code is an
improved version of Starburst99, incorporating both a new set of
evolutionary tracks from the Padova group for old and low-mass stars
\citep{vazquez05}, as well as a high-resolution (0.3 {\AA}) optical
spectral library (excluding WR stars) covering the full HRD
\citep{lucimara05}.

We have implemented optical WR spectral line synthesis into
Starburst99 by incorporating the high resolution (0.3 {\AA})
\textit{University College London} (UCL) grids of expanding, non-LTE,
line-blanketed model atmospheres for WR stars \citep*{snc02}. The UCL
models have been calculated for the five metallicities represented in
Starburst99 (0.05, 0.2, 0.4, 1 and 2 \zsun) using the \textsc{cmfgen}
code of \cite{hillier98}. The WR grids assume mass-loss scales with
metallicity ($\mdot-Z$) and replace the pure helium, unblanketed, WR
models of \citet{schmutz92}. Hereafter, we refer to the computed
models as the `SB99+UCL' models.

\subsubsection{Model parameters}

Three models were calculated for an assumed $10^6$ {\msun}
instantaneous burst between 0 and 8 Myr for a 0.1--100 {\msun} Kroupa
IMF. We choose to use the Geneva group solar and LMC metallicity
stellar evolutionary tracks with enhanced mass-loss rates
\citep{meynet94}. For the LMC metallicity tracks we use both the solar
and LMC WR model atmospheres to test the effects of $\mdot-Z$ scaling
({\ie} the default scaling with $\mdot$ is turned off). We adopt the
Geneva evolutionary tracks over those of the Padova group \citep[][and
references therein]{girardi00} due to the better treatment of
assigning WR atmospheric models to the evolutionary tracks
\citep[see][]{vazquez05}.

\subsubsection{Optical}\label{sb99_optical}

The massive star content of Tol\,89 is obtained by scaling the
best-fitting SB99+UCL models to match the continuum flux of the
dereddened spectrum over the wavelength range of 3000--7000 {\AA}.

The observed spectrum is first corrected for foreground galactic
extinction and the necessary amount of internal extinction is then
applied to the observed data to match the slope of the best-fitting
model, which is adjusted to match the optical continuum flux of the
dereddened observed data.  This scaling, along with the massive star
content calculated by Starburst99 for our best-fitting (age) model,
then gives the mass of the burst event, and thus the massive star
population. The predicted N(WR)/N(O) and N(WC)/N(WN) number ratios
calculated by Starburst99 are, of course, unchanged by this
mass-scaling. The results of the fitting procedure are given in
Table~\ref{sb99+ucl}.

\begin{table*}
\caption{Massive star content derived from SB99+UCL modelling of 
an instantaneous bursts with a Kroupa IMF ($\alpha=1.3, 2.3$) and the
following mass boundaries: 0.1 ($M_{\rm{low}}$), 0.5 and 100
($M_{\rm{up}}$) {\msun}. LMC and Solar metallicity models are presented,
in which O and WR populations result from optical
continuum and emission line fits, respectively.}
\label{sb99+ucl}
\begin{tabular}{llllllllllll}
\hline\hline
Model & \mcfive{LMC, Z$=0.4${\zsun}} &  & \mctwo{LMC, Z$=0.4${\zsun}} & & \mctwo{Solar}\vspace{1mm}\\
 & \mcfive{UV} & &  \mctwo{Optical} & & \mctwo{Optical} \\\cline{2-6}\cline{8-9}\cline{11-12}
Region & \mcone{A1} & \mcone{A2} & \mcone{A3} & \mcone{A4} & \mcone{B1} & & \mcone{A} & \mcone{B} & & \mcone{A} & \mcone{B} \\\hline

Age (Myr) & 4.5$\pm$0.5 & 5.0$\pm$1.0 & 3.0--5.5 & 3.0--5.5 & $<3.5$ & & 3.5--4.0 & $<2.5$ & &$4.0\pm0.5$ & $<2$ \\

Mass ($\times10^5$ {\msun}) & 1--2 & 0.2--0.6 & 0.04--0.17 & 0.01--0.05 & 0.3 & & 2--3 & 5 & & $4\pm1$  & 6 \\     

N(O) & 310--680 & 40--190 & 10--60 & 5--20 & 80--120 & & 660--910 & 1630--1740 & & 670--1620 & 2030--2120 \\ 

N(WN) &... &... &... &... &... & & 2--3 &... & & 5--20 &... \\ 
N(WC) &... &... &... &... &... & & 60--90 &... & & 90--240 &... \\  
N(WR)/N(O) &... &... &... &... &... & & 0.1 &... & & $\sim0.2$   &... \\
N(WC)/N(WN) &... &... &...&... &... & & 17--35 &... & & 7--31 &... \\ 

\hline
\end{tabular}
\end{table*}

\begin{figure}
\includegraphics[scale=0.73,angle=0,clip=true]
{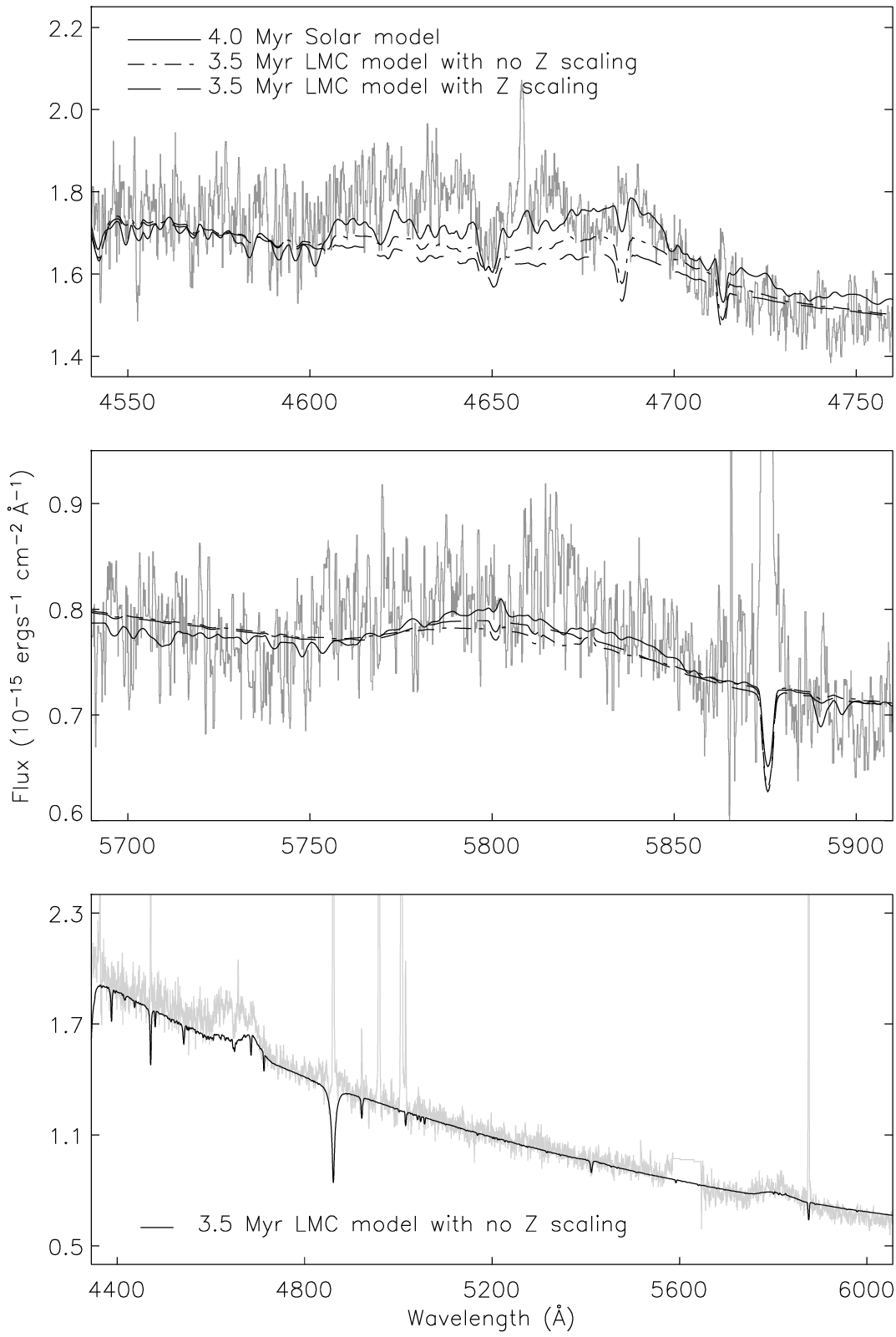}
\caption{Plots showing the best fitting (age) SB99\,+\,UCL
  models to the blue and yellow WR bumps of Tol\,89-A for solar and
  LMC metallicities. A 3.5 Myr, $Z=0.4${\zsun} metallicity
  instantaneous burst model for a 0.1--100 {\msun} Kroupa IMF is shown
  over the spectral range 4400--6000 {\AA}. The observed cluster
  spectrum has been velocity corrected and binned for
  clarity.}\label{popsynth}
\end{figure}

In Fig.~\ref{popsynth} we show the best-fitting (age) SB99+UCL
synthetic spectra to the blue bump in Tol\,89-A for each of our
models. The ages derived are slightly younger than the $\sim4.5$ Myr
derived in Section~\ref{knot_ages}, {\cf} $4.0\pm0.5$ (solar) and
3.5--4.0 Myr (LMC). We can see that the best-fitting model to the blue
WR bump, and to {\heii}4686 in particular, is obtained for a 4 Myr
solar model; although the solar model does not give the best fit to
the overall continuum shape.

While the LMC models do give a better fit to the continuum, they fail
to predict the strength of {\heii}4686, even with $\mdot-Z$ scaling
for WR stars switched off. This result is not entirely unexpected
since the initial mass for WR formation increases with decreasing
metallicity \citep{meynet94} and therefore we expect fewer WR stars to
form for a given IMF. In neither of the models is the strength of
{\nv}4620 matched, although all provide a good fit to {\civ}5808. We
note, however, that even the empirical LMC WR templates of CH06 fail
to reproduce the observed profile of the blue bump (see
Section~\ref{empirical}).

We determine the mass of the burst in knot A to be in the range
(4$\pm$1) $\times10^5$ {\msun} (Solar) and 2--3 $\times10^5$ {\msun}
(LMC). Applying the same approach to knot B, we derive ages of less
than 2.0 and 2.5 Myr, plus masses of 6 and 5 $\times10^5${\msun}, for
the Solar and LMC models, respectively. Using these masses, and the
population predictions made by Starburst99, we derive the O and WR
star numbers given in Table~\ref{sb99+ucl}. Using the LMC models we
derive N(O) = 660--910, N(WN) = 2--3 and N(WC) = 60--90 for Tol\,89-A.
The N(WR)/N(O) and N(WC)/N(WN) ratios are 0.1 and 17--35,
respectively. The large N(WC)/N(WN) ratios predicted by Starburst99
are discussed in Section~\ref{discussion}.  For knot B we derive N(O)
= 2030--2120 (Solar) or 1630--1740 (LMC) as shown in Table~\ref{sb99+ucl}.

For Tol\,89-A in particular, the O star numbers we derive from optical
continuum fits with Starburst99 are in good agreement with the numbers
indirectly derived from the nebular {\hb} line luminosity in
Section~\ref{est_no}: {\cf} $\sim690$ with $\sim660$--910 for
Tol\,89-A and $\sim2800$ with $\sim1600$--1700 for Tol\,89-B for
L(\hb) versus Starburst99 modelling. In addition, the N(WR)/N(O) ratio
inferred for Tol\,89-A is also in excellent agreement, {\cf} 0.2
versus 0.1--0.2, respectively.

\subsubsection{UV}\label{sb99_uv}

We apply the same technique to the UV continuum, scaling the
best-fitting (age) LMC Starburst99 models determined in
Section~\ref{cluster_ages} to match the flux levels of the dereddened
cluster spectra of A1--4 and B1. We derive the following masses for A1
and A2 of $\sim$1--2$\times10^5$ and 2--6$\times10^4$ {\msun}, giving
N(O) $\approx310$--680 and $\approx40$-190 respectively. For A3 and A4
we derive masses of 0.4--1.7$\times10^4$ and 1--5$\times10^3$ {\msun},
giving N(O) $\approx10$--60 and $\approx5$-20 respectively. The sum of
the masses determined for A1--4 are consistent with the value derived
from the optical, {\cf} $\sim$1--3$\times10^5$ and
$\sim2$--3$\times10^5$ {\msun} respectively. Similarly, the O star
numbers are also in good agreement: {\cf} $\sim660$--910 and
$\sim370$--950 from the optical and UV modelling respectively. These
results are in excellent agreement with the $\sim690$ O stars derived
from the nebular {\hb} approach. For B1 we derive a mass of $\sim 3
\times10^4$ {\msun}, giving N(O) $\approx80$--120. This mass is
significantly smaller than the $\sim 5 \times10^5$ {\msun} derived
from the optical UVES spectrum for Tol\,89-B, which optical STIS
imaging reveals is made up of at least 5 clusters (see leftmost inset
of Fig.~\ref{stis_vis}).

\section{Discussion}\label{discussion}
\subsection{Tol\,89 as a GHR}

Tol\,89 is conspicuous by being the only large site of star formation
in NGC\,5398. \cite{johnson03} note that Tol\,89 is among the most
luminous radio {\hii} regions yet observed and is comparable to
NGC\,5471 in M\,101.

Tol\,89 is composed of three knots of star formation which we denote
as A, B and C. Optical HST/STIS imaging resolves the two brightest
optical knots of star formation (A and B) into multiple clusters. Each
is a massive burst of star formation in its own right, having produced
multiple young compact massive clusters (R$_{\rm{eff}}\le3$ pc,
M\,$>\sim10^{3-5}$ {\msun}) over very short time scales ($\sim$ few
Myr). The brightest cluster in knot A (A1), for example, has a mass of
$\sim$1--2$\times10^{5}$ {\msun}, while a total mass of
$\sim6\times10^{5}$ is inferred for the ionizing sources at the heart
of Tol\,89-B. A further two young compact (R$_{\rm{eff}}\sim2$ pc)
massive clusters C1 and C2 are identified in the STIS UV image. The
presence of so many young compact massive clusters within such an
isolated GHR makes Tol\,89 a rather unusual object. Typical GHRs in
non-interacting, late-type spiral galaxies tend to host groupings of
fewer and less massive ($\sim10^{3-4}$ {\msun}) clusters \citep[{\eg}
NGCs\,5461, 5462 and 5471 in M\,101;][and NGCs\,592, 595 and 588 in
M\,33; \citeauthor{pellerin06} \citeyear{pellerin06}]{chen05} or
multiple OB associations \citep[{\eg} NGC\,604 in M\,33;][]{hunter96}.

The fact that Tol\,89 is located at the end of the bar in NGC\,5398
indicates it may have been formed through gas inflow in a high
pressure environment -- the conditions under which SSCs are thought to
form \citep{elmegreen97}. However, \cite{johnson03} suggest that the
weak bars found in late-type galaxies are not strong enough to
generate the required gas inflow. It is therefore not clear why so
many young compact massive clusters have formed in Tol\,89.

\citet{larsen99,larsen00} have shown that massive star clusters do
form in normal galaxies which show no obvious signs of recent
interaction -- \textit{although these typically form in isolation}. They
find that the formation of young massive clusters is favoured in
environments with active star formation and thus suggest that their
formation in starbursts or mergers may simply be extreme cases of a
more general phenomenon.  One possible explanation for the presence of
such a massive star forming region is that the parent galaxy NGC 5398
may be undergoing some form of interaction. A detailed investigation
of NGC 5398 and its environment is needed to settle this issue.

In Table~\ref{tol89_prop} we compare the properties of Tol\,89 with
the three GHRs in M\,101 and 30\,Doradus in the LMC. It can be seen
that Tol\,89 is comparable to, but no more exceptional than, the GHRs
presented in Table~\ref{tol89_prop} in terms of its {\ha}/radio
luminosities and size. The area normalised star formation rate (SFR)
of Tol\,89 is $\approx0.1$ {\msun}~yr$^{-1}$~kpc$^{-2}$ and is
comparable to the values derived for the cluster complexes in M\,51
\citep[see table 1 of][]{bastian05} and to the definition of a
starburst galaxy \citep[0.1
{\msun}~yr$^{-1}$~kpc$^{-2}$][]{kennicutt05}. The SFR rates in
Table~\ref{tol89_prop}\footnote{The value for Tol\,89 is derived from
  the {\ha} flux estimated from the {\ha} images obtained as part of
  the \textit{Spitzer Infrared Nearby Galaxies Survey} (SINGS) Legacy
  Project \citep{kennicutt03}. The observations were carried out at
  the Kitt Peak National Observatory (KPNO) 2.1 m telescope. We note
  that this value is derived from the observed ({\ie} non-extinction
  corrected) {\ha} emission flux in accordance with the results
  presented by \cite{chen05} in their table~1.} are derived using the
prescription of \citet{kennicutt98}:
$\Sigma_{SFR}$\,({\msun}~yr$^{-1}$) $=
7.9\times10^{-42}L_{\rm{\ha}}$\,(\ergss).

\begin{table*}
\caption{Comparison of Tol\,89 with NGCs\,5461, 5462 and 5471 in M\,101 and
30\,Doradus in the LMC. Values for Tol\,89 are derived from {\ha}
images obtained as part of the SINGS Legacy Project
\citep{kennicutt03}. The observations were carried out at the Kitt
Peak National Observatory (KPNO) 2.1 m telescope. $L_{\rm{6cm}}$ 
for Tol\,89 comes from \citet{johnson03}. Values for 
the three GHRs in M\,101 and 30\,Dor in the LMC are taken from table~1 
of \citet{chen05}, except the star formation rates (SFR) which were 
calculated using the prescription of \citet{kennicutt98}. 
The $L_{\rm{\ha}}$ are not extinction corrected for any the GHRs 
presented in the table.}\label{tol89_prop}
\begin{tabular}{lccccc}
\hline\hline
GHR & Tol\,89 & NGC\,5461 & NGC\,5462 & NGC\,5471 & 30\,Dor \\\hline
Angular Size (\arcsec) & $24\times18$ & $40\times25$ & $48\times33$ 
 & $17\times17$  & $1200\times1200$ \\ 
Linear Size (pc) & $1700\times1230$ & $1400\times875$ & $1680\times1150$ 
 & $600\times600$  & $290\times290$ \\ 
$L_{\rm{\ha}}$ (\ergss) & $1.9\times10^{40}$ & $2.7\times10^{40}$ 
& $1.3\times10^{40}$ & $2.2\times10^{40}$ & $3.9\times10^{39}$ \\
SFR$_{\rm{\ha}}$ ({\msun}~yr$^{-1}$~kpc$^{-2}$) & 0.07 & 0.17 
& 0.03 & 0.47 & 0.36 \\ 
 $L_{\rm{6cm}}$  &  $6.2\times10^{26}$ & $1.4\times10^{27}$ 
&  $9.1\times10^{26}$ &  $7.4\times10^{26}$ & ...  \\   
\hline
\end{tabular}

\end{table*}
\subsection{Nebular emission lines}

We first consider the origin of the nebular {\heii}4686 emission in
Tol\,89--B.  This detection is to our knowledge the first in a WR
galaxy with a metallicity greater than 0.2 {\zsun}
\citep*[\cf.\,][]{guseva00}. The detection in Tol89-A is somewhat
ambiguous -- due to the poor quality of the observations --
nevertheless, nebular {\heii} 4686 is expected to originate from
weak-lined, metal-poor early-type WN or WC stars \citep{snc02}.  As
for Tol\,89--B, the absence of WR stars implies that either the most
massive O stars are responsible for providing the {\heii} ionizing
photons, or that other (non-photoionizing) mechanisms such as
collisional shocks may be responsible for this emission.

The intrinsic flux ratio relative to {\hb}, $I(\lambda4686)/I({\hb})$,
is $\approx 1\times10^{-3}$.  \citeauthor{sv98} note that for young
bursts dominated by O stars ($t<3$ Myr), typical values for
$I$({\heii}4686)/$I$({\hb}) lie between $5\times10^{-4}$ and
$2\times10^{-3}$ (see their fig. 8). Thus, it would seem that our
$I$({\heii}4686)/$I$({\hb}) ratio is approximately consistent with a
young starburst event in the pre-WR phase (in agreement with the fact
the we do not detect any WR emissions). However, the SNC02 models --
which include a more thorough treatment of chemistry and line
blanketing in their WR models -- predict $I(\lambda4686)/I({\hb})$
ratios that are a factor of $\approx10$ lower for all ages
($\sim10^{-5}$ and $\sim10^{-4}$), in which the lowest metallicity
models predict the hardest ionizing flux distributions.  During the
pre-WR phase the SV98 and SNC02 models differ due to the use of the
\textsc{costar} (SV98) and WM-Basic (SNC02) models \citep[see][for
details]{snc02}.

Consequently, one does not anticipate nebular {\heii}4686 in young
starburst regions, such as Tol\,89-B, to arise from O stars.
\citet{garnett91} discuss {\hii} regions in which nebular {\heii}4686
is observed. In most cases, Wolf-Rayet stars provide the ionizing
source for these nebulae, except in two instances, for which an early O
star and a massive X-ray binary appear to provide the hard ionization,
the former being especially puzzling. At present, the source of
nebular {\heii}4686 is unexplained in Tol\,89-B, whilst weak-lined WN
or WC stars could provide the necessary hard ionizing photons in
Tol\,89-A.

We now consider the origin of the broad velocity components seen in
knots A and B. They occur at the same velocity which suggests that
they have a common origin, although the knot B component is narrower
compared to that of knot A (70 vs. 110 {\kms}). In addition, the broad
and narrow components appear to have a similar electron densities.
This suggests that the gas responsible for the broad component can
only be differentiated from the narrow {\hii} region component by its
width and that the two distinct components probably coexist within the
GHR. Finally, we find that the maximum velocity of the gas, as
measured by the FWZI, is 450--600 {\kms}.

Underlying broad components to nebular emission lines have been
reported in a number of studies of GHRs, and WR and starburst galaxies
\citep[see][for a review]{mendez97,homeier99}. The properties of these
components resemble those we have found in Tol\,89; they have similar
ionization conditions to the narrow components, and are spatially
extended over the star-forming knots \citep{mendez97}. The origin of
the broad component is not well understood; \citeauthor{homeier99}
consider three possible explanations: (1) it is due to integrating over
many ionized structures at different velocities; (2) it originates
from hot, turbulent gas within superbubbles created by the winds from
clusters; or (3) it is associated with some type of break-out
phenomenon such as a galactic wind. The third option is unlikely
because the broad component is not velocity-shifted with respect to
the narrow {\hii} region component. The similar densities and
ionization states of the broad and narrow components argues against
the second option.  We are then left with the first option that the
broad component is the result of integrating over shell structures and
filaments at different velocities. Studies of the well-resolved GHRs
30 Dor \citep{chu94} and NGC 604 \citep{yang96} show
that their integrated profiles have low intensity broad wings due to
fast expanding shells.  We therefore favour this explanation because
of the similar densities and ionization states of the broad and narrow
components.

The narrower width of the broad component in knot B compared to knot
A may be explained by the younger age of knot B ($<3$ Myr compared to
4.5 Myr). The cluster winds in knot A will be more advanced than
those in knot B because of the onset of WR winds and supernovae. We
would thus expect the cluster winds in knot A to have had a larger
impact on the dynamics and structure of the surrounding H II region.
This is in accord with the larger reddening we deduce for knot B, its
morphology showing a higher gas concentration (Fig. 4), as well as
the fact that we observe velocity splitting in the main nebular
component for knot A but not for knot B.

\subsection{The massive star content of Tol\,89}
\subsubsection{Clusters A1 to A4}

We first consider the difference in the massive star contents of
clusters A1 and A2.  While A2 shows a mixed population of WN and WC
stars, A1 only contains WN stars, although both have similar ages of
4.5 and 5.0 Myr. A similar difference in the populations of the two
subclusters in the bright super--star cluster SSC-A in the dwarf
starburst galaxy NGC\,1569 was discussed by \cite*{maoz01}. Their STIS
long slit optical spectroscopy revealed the presence of young WR
features ($\leq5$Myr) coexisting with an older \textit{red supergiant}
(RSG) population ($\geq4$Myr).  \citeauthor{maoz01} suggest that there
is a dichotomy in the population of the two subclusters NGC~1569--A1
and A2, despite their similar ages of 5 Myr, in which the WR feature
originates solely from NGC~1569--A2.  They conclude that NGC~1569--A1
and A2 must have either widely different IMFs or widely different
abundances, or similar, anomalously high, abundances but slightly
different ages. In fact, recent estimates of the age of NGC\,1569-A2
have been revised and it is now believed to be around 12 Myr old
\citep{anders04}, thus explaining the difference in the stellar
populations of NGC\,1569--A1 and A2.  Of course, individual clusters
can host mixed (WR and RSG) populations if their ages are in the range
4--5Myr.  Westerlund~1 in the Milky Way is an example of a massive
cluster whose content has been spatially resolved into RSG, WN and WC
populations by \citet{clark05}.

For the case of the clusters A1 and A2 in Tol\,89 we find the
difference between A1 containing only WN stars, and A2 both WN and WC
stars, can be partially explained in terms of age effects due to the
rapid evolution of WR stars on time scales of typically a few $10^5$
years \citep{mm94}.  Since the WC phase follows the WN stage for the
most massive stars, one might suspect that the slightly younger
cluster (A1) contains only WN stars, while the older cluster (A2)
contains a mixed population of WN stars and WC stars.  Unfortunately,
one expects the most massive stars to advance to the WC stage within
$\sim$3Myr, which is inconsistent with the age inferred from the -- A1
dominated -- {\hb} equivalent width in knot A. Alternatively, the age
would need to exceed 5Myr for lower initial mass stars to advance
through to the WN phase, but fail to become WC stars prior to
core-collapse, which is also somewhat in conflict with the observed
{\hb} equivalent width for knot A.  Differences in WR populations
could result from variations in IMF, with a deficit in very high mass
stars for A1, although this scenario appears to be rather contrived,
given the rapid variation in N(WC)/N(WN) with age for young starbursts
\cite{sv98}.

A similar situation would appear to arise for clusters A3 and A4 in
the sense that from their derived ages (see
Section~\ref{cluster_ages}) one would also expect WR stars to be
present. While no obvious WR signatures are detected in cluster A3,
clear {\heii} 1640 emission is observed in A4 (see
Fig.~\ref{uv_spectra}). Remarkably, in cluster A4 we have been able to
detect just three mid WN stars despite the large distance of 14.7~Mpc
to Tol\,89. The detection of such a small number of WR stars is partly
made possible due to the weak continuum flux of this cluster, whose UV
output is entirely dominated by these three WN stars. In contrast to
cluster A4, the observed low S/N and stronger continuum in A3 may be
masking any WR stars present in this cluster, thus we can not exclude
the presence of a small number WN stars. We derive a mass for A3 and
A4 of $\sim 0.4$--1.7$\times 10^4$ and $\sim1$--5$\times 10^3$ {\msun}
from UV Starburst99 modelling.  According to \citet{cervino94} and
\citet{cervino02}, below $\sim 10^{4-5}$ {\msun} the IMF is no longer
well sampled, thus affecting the integrated properties of the cluster.
In such cases stellar population synthesis models can no longer be
correctly applied \citep[see {\eg}][]{jamet04}. It is possible that
the massive stellar content anomalies of A3 and A4 are due to a
stochastic sampling of the IMF as a result of the low cluster masses.

\subsubsection{Empirical constraints}

Using template spectra of LMC WN and WC type stars \citep{ch06} we
have been able to derive consistent early WN ($\approx 95$) and WC
($\approx 35$) star numbers from both optical and UV diagnostics. In
the UV, we sum the contributions from clusters A1 and A2 in order to
make direct comparisons with the optical UVES spectra of knot A; we do
not include the three WN5--6 stars derived for cluster A4 since these
will not contribute significantly to the optical blue bump which we
attribute entirely to WN2--4 stars. Based on nebular derived O star
populations, we estimate N(WR)/N(O) $\sim$0.2 for knot A, somewhat
larger than single star evolutionary models predict at LMC metallicity
\cite{sv98}.

The UV STIS data are also presented in \cite{chandar04} who derive a
{\heii}1640 flux equivalent to $\sim130$ early-type WN stars (based on
the CH06 LMC line luminosity for WN2--4 stars) from their extraction
of Tol\,89-1 (encompassing our A1 and A2). This compares well to the
results of this work, although we note that the difference is most
likely due to the choice of extinction law used by
\citeauthor{chandar04} As noted by \cite{hc06}, the use of a standard
starburst extinction law is ideally suited to spatially unresolved
galaxies. For extragalactic stellar clusters such as Tol\,89-A1(--A3)
and NGC\,3125-A1, an LMC or SMC extinction law is more appropriate.
\cite{hc06} have shown that the WR content of NGC\,3125-1 derived from
{\heii}1640 by \citeauthor{chandar04} is strongly overestimated due to
their choice of extinction law.

The absolute WR content we derive from the optical is significantly
lower than that estimated by \citeauthor{schaerer99} by about a factor
of 3; although the WC/WN ratios obtained are in good agreement ({\cf}
$\sim0.5$ and $\sim0.6$ respectively). This discrepancy is likely due
to several factors, including different adopted internal extinctions,
slit widths ({\cf} 1\farcs6 to our 1\farcs4) and position angle (PA
39{\degr} and 90{\degr}). Note that \citeauthor{schaerer99} adopt the
global extinction value derived for Tol\,89 by \citet{terlevich91}
(0.12 mags), whilst we derive an extinction value of zero mags for
Tol\,89-A directly.

\subsubsection{Starburst99 model constraints}

In Section~\ref{sb99} we compared three SB99+UCL models to the
observed WR profiles in knot A; \textit{1)} Solar; \textit{2)} LMC
with $\mdot-Z$ for WR stars switched on; and \textit{3)} LMC with
$\mdot-Z$ for WR stars switched off. Fig.~\ref{popsynth} shows that
while we are able to obtain good fits to the yellow WR bump, the fit
to the blue bump is rather unsatisfactory -- particularly in the case
of the LMC metallicity models. The best agreement is achieved for
solar metallicity models, which we acknowledge to be unphysical in
view of the low metallicity of Tol\,89.

Using the LMC models, the predicted number of WR stars is in
reasonable agreement with the empirical results, approximately a
factor of $\sim1$--2 times smaller ({\cf} $\sim60$--90 to 130), while
the O-star numbers are also similar from the stellar continuum
($\sim660$--910) and {\hii} region ($\sim690$) analyses respectively.
The SB99+UCL spectral synthesis models predict N(WR)/N(O) $\sim0.1$
versus $\sim0.2$ from direct stellar (WR) and indirect nebular (O)
results. Unfortunately, the major failure of the spectral synthesis
approach relates to the distribution of WR stars, which is observed to
be primarily WN stars in contrast to a predicted dominant WC
population. As a consequence, the fit to the blue bump is poor since
the WN population is greatly underestimated.

However, let us recall that non-rotating evolutionary models are at
present used in Starburst99 synthesis models.  \cite{mm03,mm05} have
shown that the inclusion of rotational mixing in their evolutionary
models increases the WR lifetimes -- mainly as a result of the
increased duration of the H-rich phase -- and lowers the initial mass
limit for the formation of WR stars. The combination of these two
factors is to decrease the N(WC)/N(WN) ratio and increase the
predicted N(WR)/N(O) ratio, bringing predictions closer to our
empirical results for Tol\,89-A.  The potential use of synthetic WR
line bumps in Starburst99 models as diagnostics of WR populations is
at present severely hindered by the lack of rotational mixing in
evolutionary models. Until then, one should treat detailed N(WR)/N(O)
and WR subtype distributions predicted by such models with caution.

\section{Summary}\label{summary}

We have presented new high spectral resolution VLT/UVES spectroscopy
and archival HST/STIS imaging and spectroscopy of the giant
{\hii} region Tol\,89 in NGC\,5398. 

From optical HST/STIS imaging we resolve the two brightest optical
knots of star formation, Tol\,89-A and B, into individual, young
compact massive clusters A1--4 and B1 (R$_{\rm{eff}}\le3$ pc,
M\,$>\sim10^{3-5}$ {\msun}). We derive ages for knots A and B of
$\sim4.5$ and $<3$ Myr respectively. We determine a mass from
Starburst99 modelling for the brightest cluster in A (A1) of $\sim$
1--2$\times10^{5}$ {\msun}. A total mass of $\sim6\times10^{5}$
{\msun} is inferred for the ionizing sources at the heart of knot B. A
further two young compact massive clusters (R$_{\rm{eff}}<2$ pc; C1
and C2) are identified in the STIS UV image which may also fall into
the SSC (mass) category. In total, we identify at least seven young
massive compact clusters in the Tol 89 star-forming complex. We find
that the GHR properties of Tol 89 are similar to the three GHRs in
M101 and 30 Dor. Tol 89, however, contains six clusters of SSC
proportions whereas the other four comparison GHRs do not.  Tol 89 is
therefore exceptional in terms of its cluster content and its isolated
location in the late-type galaxy NGC 5398.

In agreement with the results of \citeauthor{schaerer99} 1999, we show
that the WR emission is localised to the region with maximum stellar
continuum, knot A; while STIS UV spectroscopy reveals that the WR
stars are confined to clusters A1, A2 and A4, with early WC stars
located only in A2. We have modelled the observed WR line profiles
using the empirical template spectra of LMC WN and WC stars presented
in \cite{ch06}, revealing $\sim95$ early WN stars and $\sim35$ WC
stars in Tol\,89-A.  WR populations inferred from our empirical
technique are consistent between optical and UV diagnostics, and so
are well constrained. For clusters A1, A2 and A4, we obtain
N(WC)/N(WN) $\sim0$, $\sim3$ and $\sim0$, respectively. It is feasible
that the slight difference in ages of clusters A1 and A2 (cf $\sim4.5$
and $\sim5$ Myr respectively) is responsible for this difference in
their massive star population, although differences in IMF cannot be
excluded. In cluster A4 we have been able to detect three mid WN stars
despite the large distance to Tol\,89, which is testament to the
dominance of WR stars at UV wavelengths. The detection of so few WN
stars may be the result of stochastic sampling of the IMF.  From
nebular {\hb} emission, we obtain N(O)$\sim690$ and 2800 for knots A
and B, from which we infer N(WR)/N(O)$\sim0.2$ for the former region.
For knot B, N(O) is a factor of two smaller than recent radio
observations by \cite{johnson03}.

We have constructed complementary Starburst99 \citep{leitherer} models
in which optical spectral synthesis of WR stars has been implemented
using the UCL WR star grids of \cite{snc02}. O star populations from
optical continuum flux distributions are in good agreement with
nebular results, although the SB99 + UCL models fail to reproduce the
observed strength of the blue WR bump, because too few WN stars are
predicted in the evolutionary models. The inclusion of evolutionary
tracks with rotational mixing \citep{mm03,mm05} should help to resolve
this issue by increasing the lifetime of WNs during the H-rich phase.
Nevertheless, the total WR populations obtained in this way agree with
the empirical results to within a factor of 2--3.

From an analysis of the optical nebular emission lines, we confirm
previous determinations that Tol 89 has an LMC-type metallicity.
Nebular {\heii} 4686 is observed in Tol\,89--B and perhaps Tol\,89--A
from UVES and STIS spectroscopy. The latter is expected for early-type
WR populations at low metallicity, whilst {\heii} 4686 emission from
starburst regions prior to the WR stage is not predicted, unless this
is formed within a shocked region surrounding young O star
populations, as may be the case for N44C in the LMC.  We detect
symmetrical broad components in the strongest nebular lines with
velocities up to 450--600 {\kms}. We find that this high velocity gas
has similar properties to the {\hii} gas as revealed by the narrow
components, and suggest that it is the result of integrating over
shell structures and filaments at different velocities within the GHR.

\section*{acknowledgements}

FS acknowledges financial support from PPARC and the Perren Fund, UCL.
PAC acknowledges financial support from the Royal Society. The authors
also wish to thank the referee for useful comments and suggestions,
Bill Vacca for useful discussions, Rupali Chandar for sending us her
spectrum of Tol\,89--1 and Nate Bastian for his help in using the {\sc
  ishape} routine. This paper is based on observations collected at
the European Southern Observatory, Chile, proposal ESO 73.B-0238(A).
This paper is also based on observations taken with the NASA/ESA {\it
  Hubble Space Telescope}\/ which is operated by the Association of
Universities for Research in Astronomy, Inc. under NASA contract
NAS5-26555.  It made use of HST data taken as part of GO program 7513
(PI C.  Leitherer). The Image Reduction and Analysis Facility ({\sc
  iraf}) is distributed by the National Optical Astronomy
Observatories which is operated by the Association of Universities for
Research in Astronomy, Inc. under cooperative agreement with the
National Science Foundation. {\sc stsdas} is the Space Telescope
Science Data Analysis System; its tasks are complementary to those in
{\sc iraf}.

\bibliographystyle{mn2e} 

\bibliography{tol89} 

\bsp 

\label{lastpage}

\end{document}